\documentclass[aps,prx,twocolumn,showpacs,superscriptaddress,groupedaddress]{revtex4-2}

\usepackage{graphicx}
\usepackage{dcolumn}
\usepackage{bm}

\usepackage[caption=false,font=footnotesize,captionskip=-3mm,farskip=-0mm]{subfig}
\usepackage{amssymb}
\usepackage{epsfig}
\usepackage{amsmath}
\usepackage{textcomp}
\usepackage{url}
\usepackage{float}
\usepackage{color}
\usepackage{cases}
\usepackage{relsize}
\usepackage[normalem]{ulem}
\usepackage{comment}

\newcommand{\black}{\textcolor{black}}

\newcommand\bluesout{\bgroup\markoverwith{\textcolor{blue}{\rule[0.4ex]{2pt}{1.0pt}}}\ULon} 
\usepackage[colorlinks=true, allcolors=blue]{hyperref}

\begin{document}
\title{Non-Markovian Rock-Paper-Scissors games}
\author{Ohad Vilk$^{a,\ast}$} 
\author{Mauro Mobilia$^{b,\dagger}$} 
\author{Michael Assaf$^{a,\times}$}
\affiliation{$^a$ Racah Institute of Physics, The Hebrew University of Jerusalem, Jerusalem 91904, Israel}
\affiliation{b Department of Applied Mathematics, School of Mathematics, University of
Leeds, Leeds LS2 9JT, U.K.}
\affiliation{$\ast$ Ohad.Vilk@mail.huji.ac.il, $\dagger$ M.Mobilia@leeds.ac.uk, $\times$ Michael.Assaf@mail.huji.ac.il}


\begin{abstract}
There is mounting evidence that species interactions often involve long-term memory, with highly-varying waiting times between  successive events and long-range temporal correlations. Accounting for memory undermines the common Markovian assumption, and dramatically impacts key ingredients of population dynamics including birth, foraging, predation, and competition processes.  
Here, we study a critical aspect of population dynamics, namely non-Markovian multi-species competition. This is done in the realm of the zero-sum rock-paper-scissors (zRPS) model that is broadly used in the life sciences to metaphorically describe  cyclic competition between three interacting species. We develop a general non-Markovian formalism for multi-species dynamics, allowing us to determine the regions of the parameter space where each species dominates. 
In particular, when the dynamics are Markovian, the waiting times are exponentially distributed and the fate of the zRPS model in large well-mixed populations is encoded in a remarkably simple condition, often referred to as the ``law of the weakest'' (LOW), stating that the species with the lowest growth rate is the most likely to prevail. We show  that the survival behavior and LOW of the zRPS model are critically affected by non-exponential waiting times, and especially, by their coefficient of variation. Our findings provide key insight into the influence of long waiting times on non-Markovian evolutionary processes. 
\end{abstract}
\maketitle

Ecosystems consist of a large number of competing  species,
and it is of paramount importance to study the mechanisms
 affecting their probability of extinction and survival.
 It
is well known that random birth and death events cause demographic
fluctuations that can ultimately lead to species
extinction or fixation – when one species takes over the entire
population. As demographic
fluctuations are
strong in small communities and weak in large populations, various
dynamics as well as survival and fixation scenarios appear in communities of different
size and structure, see, e.g., Refs.~\cite{Tainaka1993,Frean2001,Kerr2002,Ifti2003,assaf2006spectral,Reich2006,Reich2007,Szabo_2007,assaf2010extinction,AM10,Gallas10,Review2014,assaf2017,Review2018,West2018,West2020,Liao2020,giuggioli2022spatio}. For example, experiments on three-strain colicinogenic
microbial communities have demonstrated that cyclic rock-paper-scissors-like competition led to intriguing
behavior, with only the colicin-resistant strain surviving in large well-mixed populations, and to the long-time \black{coexistence}  of all species on Petri dishes~\cite{Kerr2002}. In this context,  ``rock-paper-scissors'' games have received much attention and  served as  paradigmatic models for the dynamics of species in cyclic competition, see, e.g.,~\cite{Sinervo_96,Frach96,Hofbauer,Frach98,Kerr2002,Ifti2003,Nowak,Reich2006,Reich2007,Perc_2007,Szabo_2007,Alava08,Claussen08,MM10,Gallas10,Jiang_2011,he2010,he2011,SMR13,SMR14,Kelsic2015,toupo15,MRS16,Postlethwaite17,Review2018,West2018,West2020,Liao2020,Kerr2002,Pleiming19,Pleiming20,Hens22}. 

The conditions favoring one species over other competing ones 
generally depend on numerous complex factors, e.g. the availability of nutrients or the presence of toxins in the ecosystems. 
However, the fixation and survival probabilities of the zero-sum 
rock-paper-scissors (zRPS) model, in which what one gains is exactly what the opponent loses,  satisfy a remarkably  simple relation in large well-mixed  populations~\cite{Tainaka1993,Frean2001,Berr2009}:
the species with lowest per-capita predation-reproduction rate (lowest payoff) is the most likely to survive and fixate the population.
This {\it counterintuitive} result, often referred to as the ``law of the
weakest'' (LOW)~\cite{Frean2001,Berr2009},  
becomes asymptotically a zero-one law in large populations, where the species with lowest payoff fixates the population and the others go extinct  with a probability approaching one~\cite{Tainaka1993,Frean2001,Berr2009}.
 The LOW has been studied in various settings, see, e.g., Refs.~\cite{Tainaka1993,Frean2001,Berr2009,West2018,West2020},  
 including in recent laboratory-controlled experiments.
Notably, Ref.~\cite{Liao2020}  studied cyclic competition
 of three strains of {\it E. coli} and found that 
 the `weakest strain' dominates the microbial community.
 
  The LOW has been derived when the underlying stochastic dynamics are interpreted as a Markov process, with 
 exponentially distributed
   waiting times (also referred to as interevent, holding or residence times), see, e.g., \cite{Allen}. 
 In many situations, however, species interactions may involve time delays or different time scales, often yielding  memory effects and hence the violation of the Markov assumption. In this case, the  waiting-time distribution (WTD) is no longer exponential, and this can  significantly affect the evolutionary dynamics, resulting, e.g., in correlations, amplified oscillations, or enhanced extinction probabilities~\cite{Viswanathan99,Metzler14,Wosniack17,Guinard21,vilk2022ergodicity,Jafarpour23,Costa24,Waclaw24}. 
In the context of animal behavior, optimal search strategies are often related to non-exponentially distributed  interevent times~\cite{Viswanathan99,Metzler14,Wosniack17,Guinard21,vilk2022ergodicity,vilk2022unravelling,vilk2022phase,vilk2022classification}. It has notably been reported 
that environmental variability, affecting  resource availability, 
can   result in a heavy-tailed  WTD which in turn shapes the
 population dynamics, see e.g.~\cite{Metzler14,Wosniack17}. For instance, 
heavy-tailed  WTDs  characterizing {\it Caenorhabditis elegans} dynamics, have recently been shown  to be associated with slow adaptation and to yield long-range correlations~\cite{Costa24}.

While non-exponential WTDs have been found to
lead to strong stochastic oscillations and to enhance extinction in a two-species predator-prey model~\cite{Waclaw24}, to the best of our knowledge,
the
fixation/survival behavior of zRPS games with non-Markovian
dynamics has not been studied. In  particular, it is unknown how   
the fixation properties change when the reactions have non exponential WTDs. In this work,  we systematically analyze how 
different examples of WTDs alter the LOW in the paradigmatic zRPS model, and hence shed further light on the  influence of WTDs on the evolution of  non-Markovian processes in the presence of long-term memory.


\textbf{Basic Model. }We consider a well-mixed population of constant size $N$ consisting of individuals of three species: $n_A$, $n_B$ and $n_C$ individuals of species $A$, $B$ and $C$, respectively, with 
$n_A\!+\!n_B\!+\!n_C\!=\!N$. The species are in cyclic competition: $A$ outcompetes $B$, which dominates $C$, which in turn kills and replaces $A$, closing the cycle. That is, in this general zRPS model, sometimes  referred to as cyclic Lotka-Volterra model~ \cite{Frach96,Frach98,Reich2006,Review2014,West2018,West2020},
each species is the predator of another, and the prey of the third species. Each predator-prey interaction consists of a ``predation with reproduction'' event, where the prey  is killed and simultaneously replaced by an individual of the predating species. The dynamics can thus be represented by the reactions [see Eq.~(\ref{MEexp})]:
\begin{eqnarray}
 \label{eq:reac}
 A+B  &\stackrel{k_{\rm A}/N}{  \longrightarrow} A+A \nonumber\\
  B+C  &\stackrel{k_{\rm B}/N}{  \longrightarrow} B+B \\
   C+A  &\stackrel{k_{\rm C}/N}{  \longrightarrow} C+C, \nonumber
\end{eqnarray}
where $k_{\rm A}$, $k_{\rm B}$, $k_{\rm C}$ are predator-prey interaction rates. %

\begin{figure}[t]
\centering
\subfloat[]{
  \centering
 \includegraphics[width=.40\linewidth]{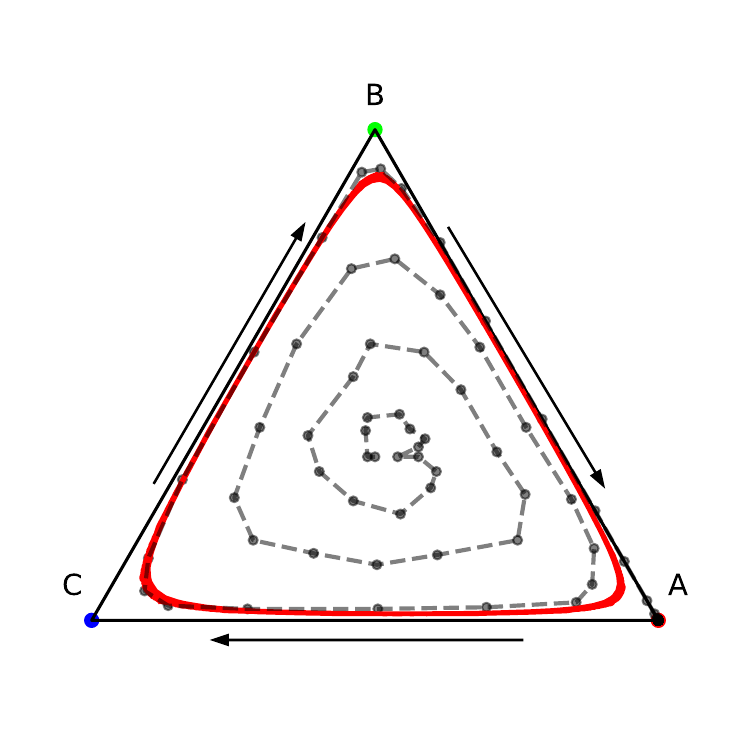}
}\hspace{-4mm}
\subfloat[]{
  \centering
 \includegraphics[width=.40\linewidth]{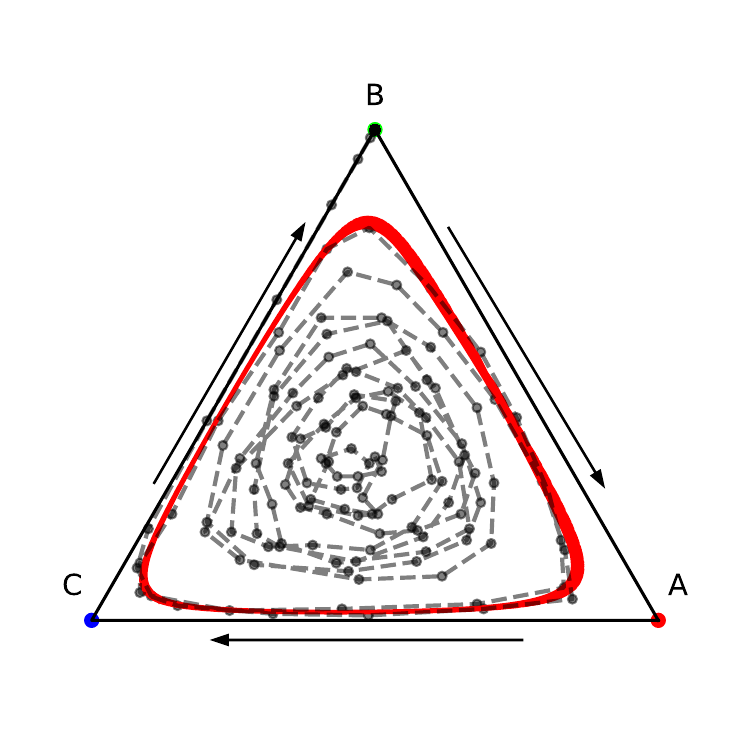}
} \vspace{-5.5mm}
\subfloat[]{
  \centering
\includegraphics[width=.40\linewidth]{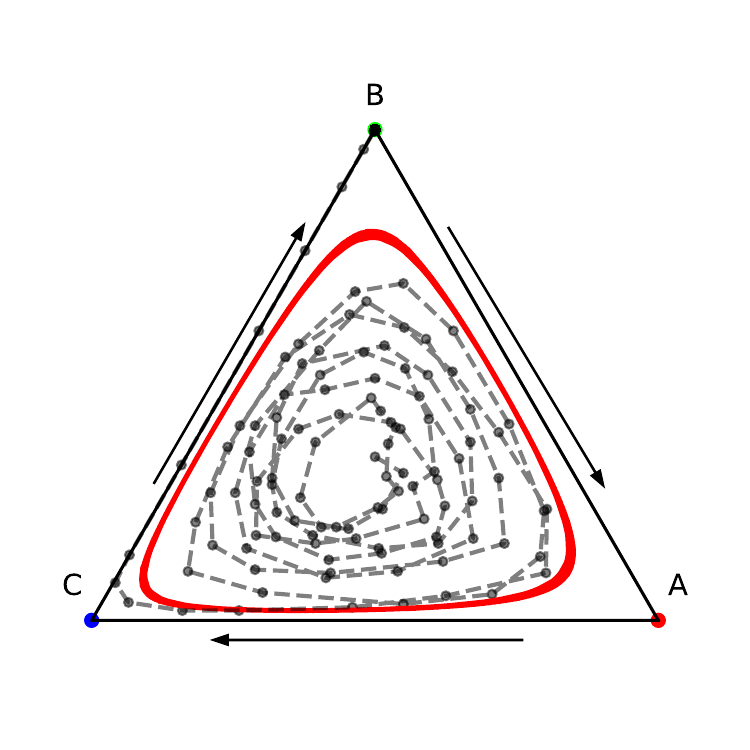}
}\hspace{-4mm}
\subfloat[]{
  \centering
 \includegraphics[width=.40\linewidth]{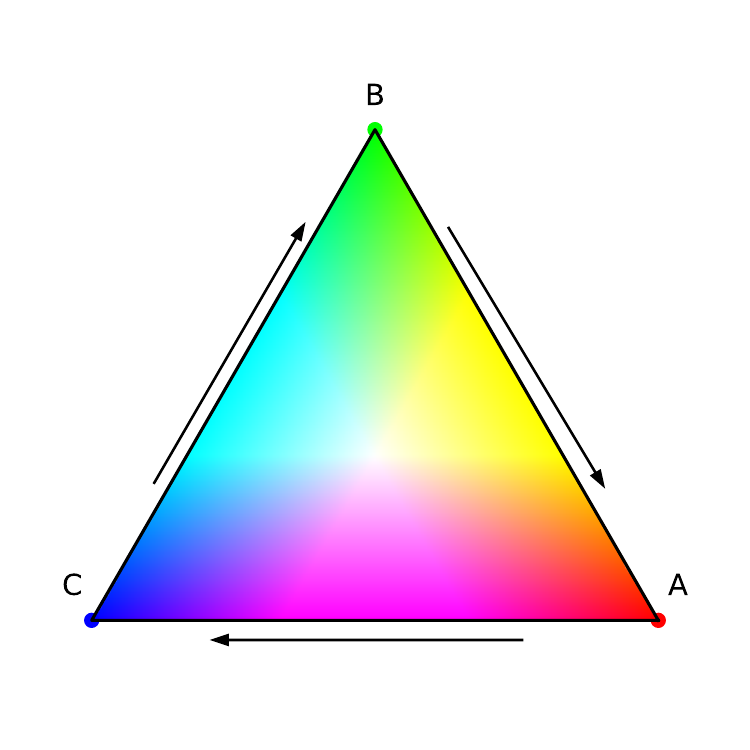}
} \vspace{-4.5mm}
\caption{(a-c)  Dynamics in the  
the ternary simplex (phase space) for the zRPS model with  exponential WTD in (a).
In (b) and (c) the last two reactions of  \eqref{eq:reac} have an exponential WTD, while the
first reaction has 
a power-law WTD~(\ref{eq:Psi-PL}) with $(k_A,\alpha_{\rm A}) = (0.8,2.5)$ in (b),  and a gamma WTD~(\ref{eq:Psi-G}) with $(k_A,\alpha_{\rm A}) = (0.8, 0.8)$ in (c). In (a-c): $k_{\rm A}=0.8, k_{\rm B}=k_{\rm C}=1$ and $N=100$.  Gray dotted lines: stochastic  trajectories (single realization,  clockwise dynamics) represent  $(n_A\!,\!n_B\!,\!n_C)/N$, with initial conditions $(1/3,1/3,1/3)$. Red thick lines:
deterministic outermost orbits, see Appendix C. Each corner  corresponds to the fixation of the labeled species. (d) RGB diagram used to color code the fixation heatmaps, \black{where the letters `A' (red), `B' (green) and `C' (blue) denote the winning species associated with each color,} see text. 
}
\label{fig1}
\end{figure}

Under Markov dynamics, in the 
mean-field (MF) limit where $N\to \infty$ and demographic fluctuations are negligible, denoting by $a =n_A/N, b=n_B/N$ and $c=n_C/N$, the respective  fractions 
of $A, B$ and $C$ in the population, the zRPS dynamics
obey the set of rate equations~\cite{Hofbauer}
\begin{equation}
 \label{eq:MF}
 \dot{a}=a(k_{\rm A} b \!-\!k_{\rm C} c)\!,\;\;\dot{b}=b(k_{\rm B} c \!-\!k_{\rm A} a)\!,\;\;\dot{c}=c(k_{\rm C} a \!-\!k_{\rm B} b)\!,
\end{equation}
where, here and henceforth, the dot denotes the time derivative.
Here, the equilibrium points are the absorbing steady states $(a,b,c)=\{(1,0,0),(0,1,0),(0,0,1)\}$
and the coexistence stationary point 
\begin{equation}
 \label{eq:sstar}
 {\bf s}^*\equiv(a^*,b^*,c^*)= (k_{\rm A}+k_{\rm B}+k_{\rm C})^{-1}(k_{\rm B}, k_{\rm C}, k_{\rm A}).
\end{equation}
The absorbing steady states correspond to each species prevailing in turn and are all saddles (unstable), whereas ${\bf s}^*$ is a marginally  stable {\it nonlinear center}. In fact,  Eqs.~(\ref{eq:MF}) admit the nontrivial constant of 
motion~\cite{Hofbauer}
\begin{equation}
 \label{eq:CoM}
 {\cal R}(t)\equiv a^{k_{\rm B}}b^{k_{\rm C}}c^{k_{\rm A}}.
\end{equation}
 With the  conservation of  ${\cal R}$, the oscillatory dynamics governed by \eqref{eq:MF} are characterized by 
 neutrally-stable closed orbits, set by ${\cal R}(t)={\cal R}(0)$, surrounding ${\bf s}^*$
in the  ternary phase space simplex, see Fig.~\ref{fig1}(a) and Appendix A. One of the aspects of this study is to analyze the robustness of the dynamics predicted by \eqref{eq:MF} when memory effects modify the forms of these rate equations, see below.

In finite populations, with $N<\infty$, the zRPS dynamics  are generally modeled as a Markov process with absorbing states \cite{Reich2006,Szabo_2007,Gardiner,VanKampen,Allen,Broom}.
 In the presence of  demographic fluctuations, stemming from randomly-occurring birth and death events, see Appendix B, ${\cal R}(t)$ is no longer conserved. Here, the  stochastic trajectories in the phase space 
 follow the MF orbits for a  transient and perform random walks between them,  before hitting a boundary and then a corner of the ternary simplex (phase space), see Fig.~\ref{fig1}(a). This results in  
 the  extinction of two species and fixation of the third~\cite{Reich2006,Review2014,Review2018,West2018}.
 The ensuing fixation/survival behavior depends crucially on the fluctuations in the  number of individuals of each species that scales as $\sqrt{N}$ (their fraction scales as $1/\sqrt{N}$). 
 In this context, there has been a great interest in analyzing the influence of
 $N$ on the species survival/fixation scenarios, see, e.g. Refs.~\cite{Tainaka1993,Frean2001,Berr2009,West2018,West2020,Liao2020}. A central question concerns the survival or, equivalently, fixation probability  
 $\phi_i$ of species $i\in \{A,B,C\}$, defined as
 \[\hspace{-1mm}
 \phi_i \equiv  \lim_{t\to \infty}\!{\rm Prob}\{n_i(t) \!=\! N | n_i(0)\}\approx \lim_{t\to \infty} {\rm Prob}\{n_i(t) \!=\! N \}.\]
 \black{In large populations, $N\gg 1$, 
  $\phi_i$
for the zRPS model with Markovian dynamics are  independent of the initial conditions (except when initially one or more species are absent, or if the system starts very close to an absorbing boundary)~\cite{Frean2001,Berr2009,West2018,West2020,Review2014}, and satisfy the LOW.
Thus, since throughout this paper we assume  $N\gg 1$,
without loss of generality we consider an equal initial number of individuals of each species, $n_i(0)=N/3$. We have also
 numerically confirmed that 
$\phi_A$, $\phi_B$ and $\phi_C$ are essentially independent of the initial condition
 when $N\gg 1$.}
   
Next, we focus on studying the influence of non-exponential WTD on $\phi_i$ in large populations, and deviations from the survival/fixation scenarios arising under Markovian dynamics that are briefly summarized below.

The mean time to extinction (MTE) $t_{\rm ext}$, the average time for two species to go extinct (with fixation of the remaining one), is also a relevant quantity that depends on $N$. For the zRPS model with Markovian dynamics, the MTE has been shown to scale linearly with $N$~\cite{Reich2006,Review2014,West2020}: $t_{\rm ext}\sim N$.
 This stems from extinction/fixation being reached after 
 ${\cal O}(N^2)$ reactions (random-walk steps in  parameter space),  each occurring on a time scale ${\cal O}(1/N)$. For the zRPS model with non-exponential WTD, we still expect $t_{\rm ext}\sim N$ whenever the underlying MF dynamics are characterized by closed orbits, see below and Fig.~\ref{fig1}(b,c). Yet, a systematic study of the MTE for  non-Markovian dynamics will be done elsewhere.

\vspace{0.2cm}
\noindent
\textbf{The law of the
weakest under Markovian dynamics. }
\black{As aforementioned,}
when reactions~(\ref{eq:reac}) have exponential WTDs (Markovian dynamics), the fixation probabilities of the zRPS games \black{satisfy} the LOW~\cite{Tainaka1993,Frean2001,Berr2009,West2018,West2020,Review2014}:
for sufficiently large $N$, typically $N\!\gtrsim\! 100$, the species $i\!\in\! \{A, B, C\}$ with the lowest rate $k_i\!\in\! \{k_{\rm A}, k_{\rm B}, k_{\rm C}\}$ is the most likely to fixate the population~\cite{Frean2001,Berr2009,Review2014}:
\begin{equation}
 \label{eq:LOW}
 \phi_i > \phi_j \quad \text{  if $k_i<k_j$ for $i \neq j\in \{A, B, C\}$.}
\end{equation}
The LOW thus identifies the species $i$ with the lowest $k_i$, dubbed the ``weakest species'',  
as the most likely to fixate/survive, with a probability
$\phi_i\leq 1$ and $0<\phi_{j}<\phi_i$\black{, independent of the initial condition}. 
Moreover, in very large
populations the LOW becomes asymptotically a zero-one law~\cite{Frean2001,Berr2009}: it predicts that the weakest species has a probability one to survive while the others go extinct. Hence, for very large $N$, we have:
\begin{align}
 \label{eq:LOW01}
 \phi_m &\to 1, \phi_n,\phi_l \to 0 \quad \text{  if $k_m<k_n, k_l$,}
\end{align}
for $(m,n,l)$ being all possible permutations of $(A,B,C)$.
If two species have the same interaction rate that is less
than the other species' rate, the LOW predicts that the latter is most likely to go extinct, with a probability approaching one when $N\gg 1$, while the former have the same probability (approaching $1/2$ when  $N\gg 1$) to fixate.
The LOW thus predicts the regions of the  parameter space in which each species is most likely to prevail~\cite{Berr2009}. For  Markovian dynamics, according to Eqs.~(\ref{eq:LOW},\ref{eq:LOW01}), the borders between these phases are given by simple linear relationships between the   $k_i$'s. 

Insight into the LOW can be gained by considering the effect of demographic fluctuations on the closed orbits of the MF dynamics~(\ref{eq:MF}). When the stochastic trajectories in the phase space reach the outermost orbit defined by ${\cal R}(t)=1/N$~\cite{Berr2009,West2018},   chance fluctuations  cause  
the extinction of two species and fixation of the remaining one. From the coexistence equilibrium ${\bf s}^*$
and expression (\ref{eq:CoM}) of  ${\cal R}$, it can be argued that 
the outermost orbit is closest to the edge leading to the fixation of the weakest species, yielding the LOW~\cite{Berr2009}. 

It is worth noting that a different scenario emerges in the Markovian zRPS model in small populations ($N\!\lesssim\! 20$):  $\phi_i$'s  \black{satisfy} the so-called ``law of stay out''~\cite{Berr2009}\black{, where $\phi_A$, $\phi_B$ and $\phi_C$ depend on the initial condition~\cite{Berr2009,West2018,West2020}~\footnote{When the initial number of individuals  of each species is identical, the law of stay out predicts that  the species that is ``least engaged'' in interactions is the most likely to survive/fixate~\cite{Berr2009,West2018,West2020}.}}. Yet, here we consider large enough systems ($N\geq 10^2$) to disregard possible effects of the law of stay out.

\vspace{0.2cm}
\noindent\textbf{RPS under exponential WTD.}
The LOW of the zRPS model has been amply studied under Markov dynamics. Here,  the   rates $k_i$ of reactions~(\ref{eq:reac})
are directly related to the mean of the {\it exponential} WTD, $\psi(\tau_i)$, between two reactions~\cite{Allen}, where $\tau_i$ is the  time between two successive events in which the predating species $i\in\{A,B,C\}$ kills and replaces a prey.

Generally, in a continuous-time Markov process, waiting times are distributed according to a one-parameter  exponential
function that can be written as
\begin{equation}
 \label{eq:Psi-exp}
 \psi_{\rm ex}(\tau)=\lambda  e^{-\lambda \tau},\quad \langle \tau\rangle=\int_0^{\infty} \tau \psi_{\rm ex}(\tau) d\tau=\lambda^{-1},
\end{equation}
with the single parameter $\lambda$ coinciding with the inverse of  $\langle \tau\rangle$, the mean time separating two successive events (reactions). In the zRPS model under Markov dynamics one has $\lambda_{\rm A}=Nk_{\rm A}ab$, $\lambda_{\rm B}=Nk_{\rm B}bc$ and $\lambda_{\rm C}=Nk_{\rm C}ac$, for the reactions in Eq.~(\ref{eq:reac}). For $ \psi_{\rm ex}(\tau)$, the variance, coefficient of variation (CV, ratio of the standard deviation to the mean), and median  are respectively ${\rm var}(\tau)=\lambda^{-2}$, ${\rm CV}(\tau)=1$, and $ \bar{\tau}=\ln{2}/\lambda$.
%

Here, we are interested in  how the LOW is affected by WTDs that are {\it not} exponential, i.e. when the resulting zRPS dynamics are {\it non-Markovian} and include long-term memory. For 
simplicity our analytical derivation focuses on the case where the second and third reactions $B+C\to B+B$ and $C+A\to C+C$ are Markovian, with exponential WTDs, $\psi_{i}(\tau)=\lambda_i e^{-\lambda_i \tau}$ for $i=B,C$, whereas the first reaction $A+B\to A+A$ is non-Markovian and has a non-exponential WTD, denoted by $\psi_{\rm A}(\tau)$. 
We consider two representative choices of a power-law and gamma WTDs~\cite{Jafarpour23} with a finite mean. The former allows us to study the influence of WTD with ``heavy tails'', commonly observed in ecology and biology~\cite{Viswanathan99,Wosniack17,Guinard21,vilk2022ergodicity,vilk2022unravelling,vilk2022phase,vilk2022classification,Costa24}, and the latter allows us to investigate the role of the WTD shape (skewness, mode) on non-Markovian dynamics. Notably, we focus on the regime where the CV of the WTD is larger than that of an exponential, i.e., CV$(\tau)>1$, where large deviations from the LOW are expected.  The non-Markovian RPS processes
considered in this work are 
simulated following the method described in Appendix C. 

\vspace{0.5cm}
\noindent\textbf{\Large{Results}}\vspace{0.1cm}\\
\noindent\textbf{RPS survival behavior with power-law  WTD.}
Here, we  assume that the interevent time $\tau_{\rm A}$
of 
 the reaction $A+B\to A+A$
is distributed according to the two-parameter $(\Lambda_{\rm A},\alpha_{\rm A})$ power-law WTD:
\begin{equation}
 \label{eq:Psi-PL}
 \psi_{\rm A}(\tau_{\rm A})=\Lambda_{\rm A}~\frac{\alpha_{\rm A}}{(1+\Lambda_{\rm A} \tau_{\rm A})^{\alpha_{\rm A}+1}}, \quad \alpha_{\rm A}>1, \black{\Lambda_{\rm A}>0,}
\end{equation}
whose mean, variance, and median  are respectively 
\begin{equation}
 \label{eq:mean-pl}
 \hspace{-3mm}\langle \tau_{\rm A} \rangle\!=\! \frac{1}{\Lambda_{\rm A}(\alpha_{\rm A}\!-\!1)}, \;\;{\rm var}(\tau_{\rm A})\!=\!\frac{\alpha_{\rm A}\langle \tau_{\rm A} \rangle^2}{(\alpha_{\rm A} \!-\!2)},\;\; \bar{\tau}_{\rm A}\!=\!\frac{2^{1/\alpha_{\rm A}}\!-\!1}{\Lambda_{\rm A} },
\end{equation}
while ${\rm CV}_{\rm A}\equiv \sqrt{{\rm var}(\tau_{\rm A})}/\langle \tau_{\rm A} \rangle =\sqrt{\alpha_{\rm A}/(\alpha_{\rm A}-2)}$. While the variance and 
${\rm CV}_{\rm A}$ are finite when $\alpha_{\rm A}>2$, one can still simulate the dynamics when $\alpha_{\rm A}\leq 2$, see below. 

The natural choice  to
 directly compare the dynamics with a non-exponential WTD and its Markovian counterpart (with exponential WTDs), is to require the WTD's mean $\langle \tau_{\rm A} \rangle$ 
to match the mean waiting time under Markovian dynamics $\lambda_{\rm A}^{-1}$~\footnote{Alternatively, one can demand that the WTD's \textit{median}, $\bar{\tau}_{\rm A}$ be equal that of the exponential distribution.
This enables comparing, e.g., WTDs that have a diverging mean.}, where $\lambda_{\rm A} = N k_{\rm A} ab$. This yields
\begin{equation} \label{lamPLmean}
    \Lambda_{\rm A}  = \lambda_{\rm A}/(\alpha_{\rm A}-1).
\end{equation}
Henceforth, we assume that \eqref{lamPLmean} holds focusing on the $\alpha_{\rm A}\!\!>\!\!1$ regime (finite mean), and discuss our results chiefly in terms of the parameters $k_{\rm A}$ and $\alpha_A$.

Notably, under Eq.~(\ref{lamPLmean}), while the mean interevent time of the reaction  $A+B\to A+A$ is the same as in the Markovian (exponential) case, the variance of ${\tau}_{\rm A}$ with the power-law WTD is larger for any $\alpha_{\rm A}>1$ (for $1<\alpha_{\rm A}\leq 2$ the variance and  ${\rm CV}_{\rm A}$ of \eqref{eq:Psi-PL} diverge), see Eq.~(\ref{eq:mean-pl}). 
In fact, we notice that for the  WTD~(\ref{eq:Psi-PL}),  ${\rm CV_{\rm A}}\to 1$ (as for an exponential WTD) as $\alpha_{\rm A}\to \infty$, and ${\rm CV_{\rm A}}\to \infty$ when $\alpha_{\rm A}\to 2$. We thus expect the main differences from the exponentially-distributed case to arise when $\alpha_{\rm A}\gtrsim 1$, whereas we recover the LOW scenarios when $\alpha_{\rm A}\to \infty$.

\begin{figure}[t!]
\centering
\includegraphics[width=0.71\linewidth]{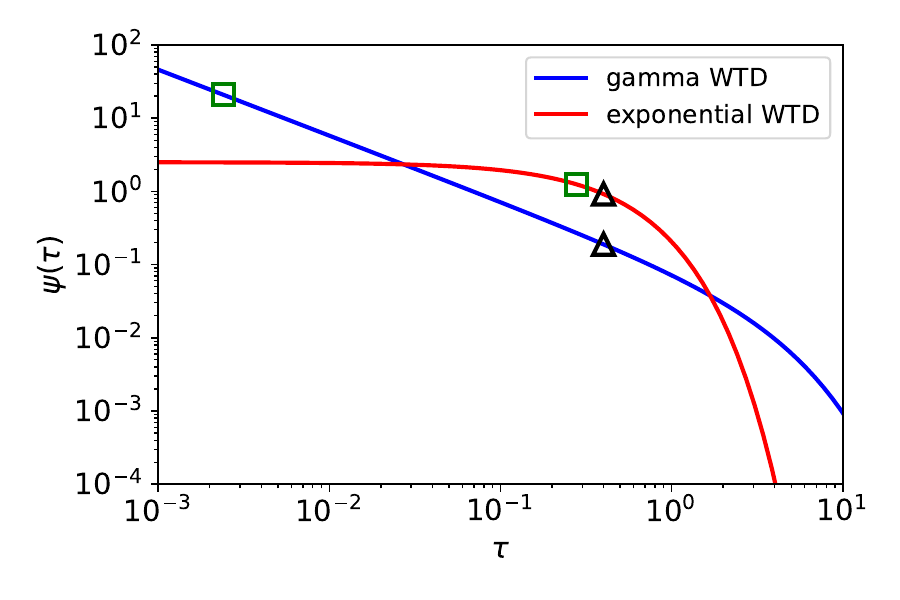}
\vspace{-7mm}
\caption{An illustration of a gamma distribution (blue line) for $\alpha_{\rm A}=0.1$ and $\Lambda_{\rm A}=0.25$, such that the mean (black triangle) equals $0.4$. The red line depicts an exponential distribution with the same mean. In contrast, the medians (green squares)  differ significantly: $0.277$ (exponential WTD) and $0.0024$ (gamma WTD).}
\label{fig2}
\end{figure}

Importantly, as shown below, for a zRPS model with a heavy-tailed WTD, the
survival/fixation behavior is not fully captured by the 
LOW as it cannot be solely inferred from the  mean interevent times of the reactions~(\ref{eq:reac}).
Intuitively, this stems from the fact that the mean time for a  reaction to occur, related to the reaction rate, is not necessarily a good measure for typical events. In fact,  while the mean time may be large, corresponding to a small reaction rate, the \textit{typical} interevent times can actually be short, see Fig.~\ref{fig2} for an illustration with a gamma WTD [see Eq.~(\ref{eq:Psi-G})] for the reaction $A+B\to A+A$ (with the others being Markovian). In this case the \textit{typical reaction rate} is larger than its mean and the LOW prediction does not generally capture the  survival/fixation scenario: i.e., even if 
$k_A<k_B,k_C$, $A$ may not be the most likely to 
survive. Notably, while this behavior can be expected for monotone-decreasing WTDs (e.g., power-law or gamma WTD with $\alpha\leq 1$, see below), it is not intuitively clear how the LOW changes for non-monotone WTDs.

\vspace{0.2cm}\noindent\textbf{Generalized rate equations under power-law WTD.}
We now consider explicitly the case
where  the reaction $A+B \to A+A$ has an interevent time distribution, given by
the power-law WTD~(\ref{eq:Psi-PL}),
with the other two reactions of \eqref{eq:reac} having exponential WTDs.
Analytical progress can be made 
 using  the formalism of continuous-time random walks~\cite{aquino2017chemical,vilk2024non},
 which leads to replace \eqref{eq:MF} by the following generalized MF rate equations:
\begin{eqnarray}\label{REPL}
&&\dot{a} =a\,b \,k_{\rm A}\Theta(a, b, c)- a\,c\, k_{\rm C},\nonumber\\&&\dot{b} = b\, c \,k_{\rm B} - a\,b\, k_{\rm A}\Theta(a, b, c),\\
&&\dot{c} = a\, c \,k_{\rm C} - b\, c \,k_{\rm B},\nonumber
\end{eqnarray}
where $\Theta(a,b,c)=\Theta_{\text{PL}}(a,b,c)$ is the memory kernel in the power-law case, see Appendix B:
\begin{equation}\label{MKPL}
    \hspace{-2mm}\Theta_{\text{PL}}(a, b, c) \!=\! \chi\left\{\!\left[1\!-\!e^{(\alpha_{\rm A}\!-\!1)\chi}\alpha_{\rm A} E_{\alpha_{\rm A}\!+\!1}[(\alpha_{\rm A}\!\!-\!\!1)\chi]\right]^{-1}\!\!\!\!-\!1\! \right\}\!.
\end{equation}
Here $\chi\!\equiv\! c(bk_{\rm B}+ak_{\rm C})/(a b k_{\rm A})$, $E_m(z) \equiv \int_1^{\infty}e^{-z\tau }\tau^{-m} d\tau\,$ is the exponential integral function, and we have set $\Lambda_{\rm A}\!=\!\lambda_{\rm A}/(\alpha\!-\!1)$ with $\lambda_{\rm A}\!=\!Nk_{\rm A}ab$. Thus, the mean interevent time of \eqref{eq:Psi-PL} equals that of an exponential WTD.

In fact, the generalized rate equations (\ref{REPL}) can be used to find the coexistence equilibrium of the zRPS model with different non-exponential  WTDs (see also the next section), and study the deviations that they cause to \eqref{eq:sstar}. When Eqs.~(\ref{REPL}) lead to closed orbits in the phase space, we can proceed as under Markovian dynamics, and infer from 
the location of the coexistence equilibrium and  outermost orbit,
which species is the most likely to fixate/survive, see Fig.~\ref{fig1}(b,c) and below.

\begin{figure}[t!]
\centering
\includegraphics[width=0.80\linewidth]{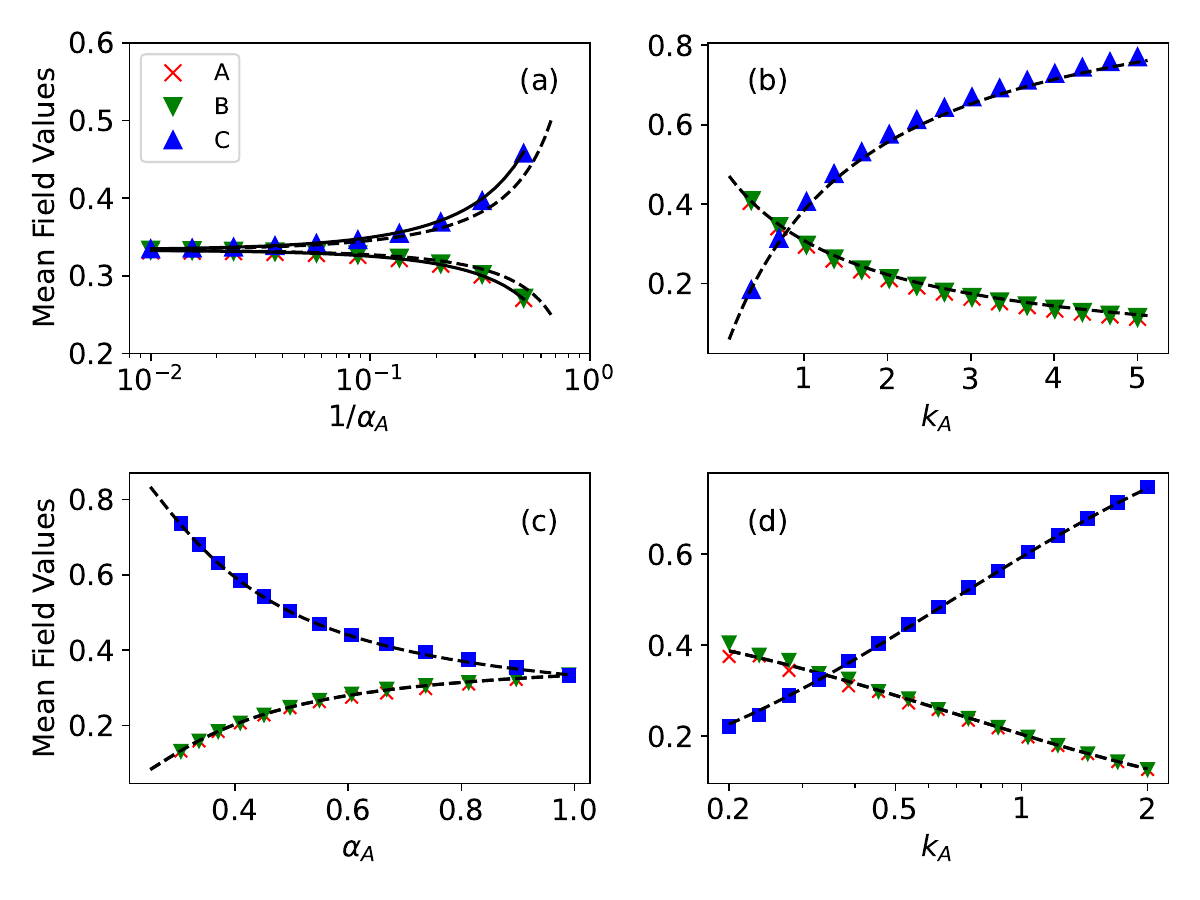}
\vspace{-5mm}\caption{Mean field steady-state concentrations of $A$, $B$ and $C$ versus $\alpha_{\rm A}^{-1}$ and $\alpha_{\rm A}$ (a,c) and $k_{\rm A}$ (b,d). In (a-b) and (c-d) the first reaction of (\ref{eq:reac}) has a power-law and gamma WTD, respectively, while the second and third reactions have exponential WTDs. 
Markers are mean field values (see legend) as obtained by averaging over stochastic simulations (see details in Appendix C). In all panels dashed lines show the analytical results: Eq.~(\ref{mean-field-PL}) in (a,b) and Eq.~(\ref{coex-gamma-full}) in (c,d). The solid lines in (a,b) show the exact expression from numerically solving Eqs.~(\ref{REPL}) for $\dot{a} = \dot{b}=\dot{c}=0$. Parameters are:  $k_{\rm A}=k_{\rm B}=k_{\rm C}=1$ (a), $\alpha_{\rm A} = 3$ and $k_{\rm B}=k_{\rm C}=1$ (b), $k_{\rm A}=k_{\rm B}=k_{\rm C}=1$ (c), and  $\alpha_{\rm A} = 0.4$ and $k_{\rm B}=k_{\rm C}=1$ (d).
}
\label{fig3}
\end{figure}

While rate equations~(\ref{REPL}) with (\ref{MKPL}) cannot be  solved analytically, a numerical solution for its coexistence stationary state $(a^*,b^*,c^*)$, with $\lambda_{\rm A}$  given by Eq.~(\ref{lamPLmean}), is shown in 
Fig.~\ref{fig3}(a,b). Figure~\ref{fig3}(a) shows $(a^*,b^*,c^*)$ versus $\alpha_{\rm A}$,
when $k_{\rm A}=k_{\rm B}=k_{\rm C}=1$ and the power-law WTD 
[Eq.~(\ref{eq:Psi-PL})]  has the same  average 
$\langle \tau_{\rm A} \rangle=1/\lambda_{\rm A}$ as under Markovian dynamics. As $\alpha_{\rm A}$ increases, we recover the well-known Markovian result, with $a^*=b^*=c^*=1/3$ for $\alpha_{\rm A}\to \infty$ [see Eq.~(\ref{eq:sstar})], whereas $a^*=b^*< c^*$ when $\alpha_{\rm A}$ is finite. Figure~\ref{fig3}(b) shows the dependence of the coexistence state on $k_{\rm A}$ for fixed $\alpha_{\rm A}$.

The limit  $\alpha_{\rm A}\gg 1$ is  particularly interesting as it is amenable to further analytical progress, and
we aim at deriving the first subleading correction to  (\ref{eq:sstar})
as a function of  $\alpha_{\rm A}^{-1}\ll 1$. To do so, we first approximate the exponential integral function $E_m(z)$ in the limit of $m,z\gg 1$, which yields:
$E_{m}(z)\!=\!\int_1^{\infty}\!e^{-z\ell}\ell^{-m} d\ell\simeq e^{-z}/(m+z)$~\footnote{This approximation holds, since the integrand is maximal at $\ell=1$, and when $m,z \gg 1$, one can Taylor-expand the integrand around $\ell=1$.}. Using this approximation and the definition of $\chi$ [below Eq.~(\ref{MKPL})], in the limit of $\alpha_{\rm A}\gg 1$ the memory kernel [Eq.~(\ref{MKPL})] becomes:
\begin{eqnarray}
\label{eq:thetaPLpprox}
 \hspace{-7mm}  \Theta_{\text{PL}}(a,b,c)\simeq \alpha_{\rm A}\chi\left[1+(\alpha_{\rm A}-1)\chi\right]^{-1},
\end{eqnarray}
where $\chi$ is a function of $a$, $b$ and $c$, given below Eq.~(\ref{MKPL}). This result is formally valid for $\alpha_{\rm A}\gg 1$, but remains rather accurate also for $\alpha_{\rm A}\gtrsim 2$, see Fig.~\ref{fig3}(a,b). As can be seen, at $\alpha_{\rm A}\to \infty$ we recover the well-known  form of $\Theta=1$, such that the rate equations (\ref{eq:MF}) of an exponential WTD are recovered, see Eqs.~(\ref{REPL}),  yielding the coexistence fixed point~(\ref{eq:sstar}). However, at finite $\alpha_{\rm A}$, we find a nontrivial correction stemming from the power-law WTD of $\tau_{\rm A}$ and non-Markovian nature of dynamics, see Fig.~\ref{fig3}(a,b). 

Focusing on $\alpha_{\rm A}\gg 1$, here we analyze the interesting case where
the  predator-prey reaction rates obey the ratio
$k_{\rm A}:k_{\rm B}:k_{\rm C}=k:1:1$, e.g.,
$k_{\rm B}=k_{\rm C}=1, k_{\rm A}=k$.
(The coexistence equilibrium point for arbitrary $k_{\rm A}$, $k_{\rm B}$ and $k_{\rm C}$ is determined in
in Appendix D.) 
Using Eq.~(\ref{eq:thetaPLpprox}), the coexistence equilibrium in this case 
reads 
\begin{eqnarray}
\label{mean-field-PL}
\hspace{-3mm}\{a^*,b^*,c^*\}=\frac{\{2(\alpha_{\rm A}-1),2(\alpha_{\rm A}-1),k(2\alpha_{\rm A}-1)\}}{4(\alpha_{\rm A}-1)+k(2\alpha_{\rm A}-1)}.
\end{eqnarray}
This shows that a power-law WTD for $\tau_{\rm A}$
generally 
changes the long-time zRPS dynamics.
In particular, the tie predicted by Eq.~(\ref{eq:sstar}) when $k=1$ is broken (even at large $\alpha_{\rm A}$), with now
 $c^*=(1/3)[1-2/(3(2\alpha_{\rm A}-1))]^{-1}$ and $a^*=b^*=(1/3)[1+1/(6(\alpha_{\rm A}-1))]^{-1}$, implying $c^*>a^*,b^*$ as found in Fig.~\ref{fig3}(a,b). 
Notably, when  $\alpha_{\rm A}$ is not too close to 1, we have numerically verified that this coexistence equilibrium is a nonlinear center, and 
species extinction/fixation thus occurs from its outermost orbit, according to the scenario outlined in the previous section, see also Appendix C. 
In Fig.~\ref{fig3}(a,b) the numerical solution of the stationary \eqref{REPL}
agrees well with the analytical approximation (\ref{mean-field-PL}) for  $a^\star, b^\star, c^\star$ and simulation results averaged over many stochastic realizations, see also Fig.~\ref{fig7} in   Appendix C. In particular,  Eq.~(\ref{mean-field-PL})
is quite accurate already when
 $\alpha_{\rm A}\gtrsim 2$.

\begin{figure}[t!]
\centering
\includegraphics[width=.98\linewidth]{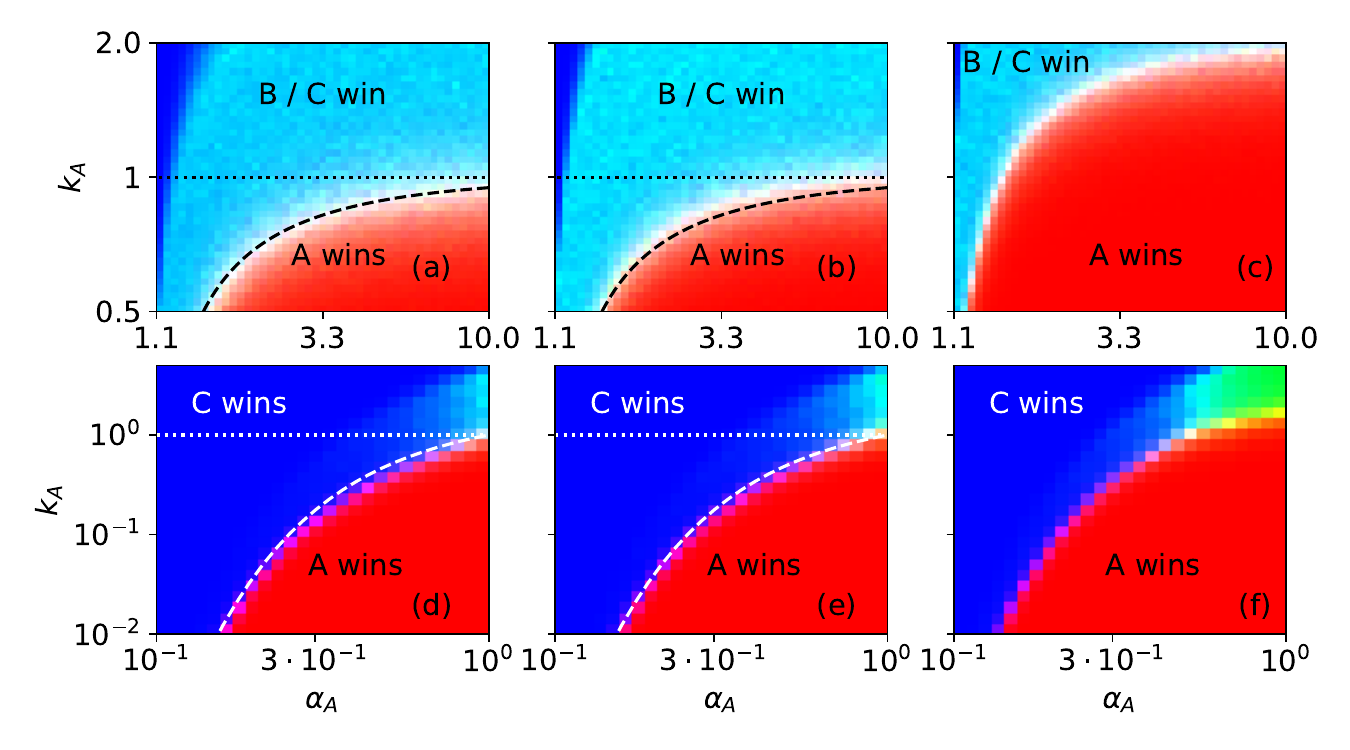}
\vspace{-5mm}
\caption{RGB fixation heatmaps for  power-law WTD~(\ref{eq:Psi-PL}) (a-c) and gamma WTD~(\ref{eq:Psi-G}) (d-f) versus $\alpha_{\rm A}$ and $k_{\rm A}$, for $N = 999$ (a-c) and $N=300$ (d-f) and $k_{\rm B}=k_{\rm C}=1$. \black{Here, red, green and  blue denote the fixation of $A$, $B$ and $C$ while cyan denotes the  fixation of $B$ and $C$ with (approximately) equal probability, see diagram of Fig.~\ref{fig1}(d).} 
In (a) and (d) the WTDs for the second and third reactions of~(\ref{eq:reac}) are exponential. In (b-c) the second and third reactions of~(\ref{eq:reac}) have power-law WTDs
with $\alpha_{\rm B} \!=\!\alpha_{\rm C}\!=\! 10$ (b) and  $\alpha_{\rm B} \!=\!\alpha_{\rm C}\!=\! 1.5$  (c). In (e-f) the second and third reactions of~(\ref{eq:reac}) have gamma WTDs with  $\alpha_{\rm B}=\alpha_{\rm C}= 0.9$ (e) and $\alpha_{\rm B}=\alpha_{\rm C} = 0.5$ (f). 
In (a-b) and (d-e) we compare our results to the theoretical curve for $k^*(\alpha)$ (dashed lines): Eq.~(\ref{kstar-PL}) for (a-b) and Eq.~(\ref{kstar-full}) for (d-e); the dotted lines denote $k_{\rm A}=1$.
}
\label{fig4}
\end{figure}

From Eq.~\eqref{mean-field-PL} we can derive a useful expression for the critical value $k^*=k(\alpha_{\rm A})$ for  which $a^*(k^*)=b^*(k^*)=c^*(k^*)=1/3$. This can be found by demanding that $a^*=b^*=c^*$ in Eq.~(\ref{mean-field-PL}), which for $\alpha_{\rm A}\gg 1$, yields:
\begin{align}
 \label{kstar-PL}
k^*(\alpha_{\rm A})\simeq 1-(2\alpha_{\rm A}-1)^{-1}.
 \end{align}
This critical value separates the  $k-\alpha_{\rm A}$ parameter space in two regions:  $c^*>a^*=b^*$ where $k>k^*$ 
and  $a^*=b^*>c^*$ where $k<k^*$.
For sufficiently large $\alpha_{\rm A}$ 
this informs on the location of the outermost orbit
of \eqref{REPL},  see also Appendices A and C,
implying  that species $A$ is the most likely to go extinct when $k>k^*$, 
while species $A$ is the most likely to fixate the population where $k<k^*$. 
This is discussed below and remarkably demonstrated in Fig.~\ref{fig4}.

\vspace{0.2cm}\noindent\textbf{RPS survival behavior with Gamma  WTD. }
We now  consider a different non-Markovian scenario where the distribution of interevent times $\tau_{\rm A}$  of 
 $A+B\!\to\! A+A$
is a two-parameter $(\Lambda_{\rm A},\alpha_{\rm A})$  gamma distribution:
\begin{equation}
 \label{eq:Psi-G}
 \hspace{-4mm}\psi_{\rm A}(\tau_{\rm A})=\frac{\Lambda_{\rm A}^{\alpha_{\rm A}}}{\Gamma(\alpha_{\rm A})}~\tau_{\rm A}^{\alpha_{\rm A}-1}e^{-\Lambda_{\rm A}\tau_{\rm A}}, \quad \text{with $\alpha_{\rm A},\Lambda_{\rm A}>0$}
\end{equation}
such that $\psi_{\rm A}$ is normalizable, with mean and variance
\begin{eqnarray}
 \label{eq:mean-G}
 \langle \tau_{\rm A} \rangle= \alpha_{\rm A}/\Lambda_{\rm A}, \quad{\rm var}(\tau_{\rm A})=\alpha_{\rm A}/\Lambda_{\rm A}^2, 
\end{eqnarray}
and ${\rm CV_{A}}(\tau)\!=\!\alpha_{\rm A}^{-1/2}\!$. 
The median does not admit a closed form but can be computed numerically for any $(\Lambda_{\rm A},\alpha_{\rm A})$. 
To directly compare  the dynamics under gamma and exponential WTDs, we demand that the mean of the gamma WTD be $\langle \tau_{\rm A}\rangle=1/\lambda_{\rm A}=1/(N k_{\rm A} ab)$. This  yields 
\begin{equation} \label{lamgammamean}
    \Lambda_{\rm A}  = \lambda_{\rm A} \alpha_{\rm A}.
\end{equation}
%
For the gamma WTD, ${\rm CV_{\rm A}}\to 1$ when $\alpha_{\rm A}\to 1$ whereas ${\rm CV_{\rm A}}\to 0$ when $\alpha_{\rm A}\to \infty$, and ${\rm CV_{\rm A}}\to \infty$ when $\alpha_{\rm A}\to 0$. We thus expect to essentially recover the LOW scenarios when $\alpha_{\rm A}\to 1$, and to find strong deviations from it when $\alpha_{\rm A}\to 0$. In the following, we focus on the regime of $\alpha_{\rm A} \leq 1$ for which ${\rm CV}_{\rm A} \geq 1$. Yet, the theory presented below is also applicable for $\alpha_{\rm A} > 1$, but a detailed treatment requires specific computational techniques that will be presented elsewhere (see Appendix C).

\vspace{0.2cm}\noindent\textbf{Generalized rate equations under gamma WTD.}
For a gamma WTD~\eqref{eq:Psi-G} of the first reaction of \eqref{eq:reac}, proceeding as above, the generalized MF rate equations 
are given by~(\ref{REPL}), with the memory kernel (see
Appendix B)
\begin{equation}\label{memory_gamma}
    \Theta_{\rm{G}}(a, b, c) = \chi\left[\left(1+\chi/\alpha_{\rm A}\right)^{\alpha_{\rm A}}-1\right]^{-1},
 \end{equation}
where again $\chi\equiv c(bk_{\rm B}+ak_{\rm C})/(a b k_{\rm A})$, and we have assumed  $\Lambda_{\rm A}=\lambda_{\rm A}\alpha_{\rm A}$. 
When $\alpha_{\rm A} = 1$, we recover $\Theta_{\rm{G}}=1$,
yielding the MF Markovian dynamics.

In this case, we can solve for the steady state of  Eqs.~(\ref{REPL}) exactly. 
Setting $\dot{a}=\dot{b}=\dot{c}=0$ in~(\ref{REPL}), we find a relation for the coexistence equilibrium: $a^*b^*k_{\rm A}\Theta_{\rm{G}}(a^*,b^*,c^*)=k_{\rm C} a^*c^*=k_{\rm B} b^*c^*$.
Here, for concreteness, we focus again on $k_{\rm A}:k_{\rm B}:k_{\rm C}=k:1:1$ (see Appendix D for the general case). Together with the relations $c^*=1-a^*-b^*$ and memory kernel~(\ref{memory_gamma}), the coexistence equilibrium here becomes
\begin{equation}
\label{coex-gamma-full}
\{a^*,b^*,c^*\}=\frac{\{2,2,k\alpha_{\rm A}(3^{1/\alpha_{\rm A}}-1)\}}{4+k\alpha_{\rm A}(3^{1/\alpha_{\rm A}}-1)}.
\end{equation}
Thus,  $a^*$ is a decreasing function of $k$ at fixed $\alpha_{\rm A}$. In addition, for $k=1$, we have $a^*=b^*>1/3$ and $c^*<1/3$ when $\alpha_{\rm A}>1$, while $a^*=b^*<1/3$ and $c^*>1/3$ when $\alpha_{\rm A}>1$, see Fig.~\ref{fig3}(c,d).  At $\alpha_{\rm A}\ll 1$, the fixed point becomes  $a^* =b^*\approx (2/k\alpha_{\rm A})
 3^{-1/\alpha_{\rm A}}$
 and $c^* \approx1-(4/k\alpha_{\rm A})
 3^{-1/\alpha_{\rm A}}$; i.e., $a^*$ and $b^*$ are exponentially
small, while $c^*$ approaches $1$ exponentially.

A simple expression for the critical value $k^*=k(\alpha_{\rm A})$ for  which $a^*(k^*)=b^*(k^*)=c^*(k^*)=1/3$ is easily found by solving $\alpha_{\rm A}k^*(3^{1/\alpha_{\rm A}}-1)=2$, yielding 
\begin{align}
 \label{kstar-full}
k^*(\alpha_{\rm A})=2\left[\alpha_{\rm A}(3^{1/\alpha_{\rm A}}-1)\right]^{-1}.
 \end{align}
This critical value separates the  $k-\alpha_{\rm A}$ parameter space in two regions, one in which $a^*=b^*>c^*$ (where $k<k^*$) and another where $a^*\!=\!b^*\!<\!c^*$ (where $k\!>\!k^*$), see below.

\vspace{0.2cm}\noindent\textbf{Fixation heatmaps for the power-law WTD.}
A systematic way to visualize  the influence of a heavy-tailed WTD on the RPS fixation behavior is by means of 
fixation heatmaps shown in  Fig.~\ref{fig4}. These 
are   RGB-coded according to the diagram of Fig.~\ref{fig1}(d) and
report the  
triplet $(\phi_{\rm A},\phi_{\rm B},\phi_{\rm C})$ versus $\alpha_{\rm A}$ and $k_{\rm A}$ -- the mean rate (per $A$-$B$  pair) of the first reaction of \eqref{eq:reac}.
According to Fig.~\ref{fig1}(d),
the phase dominated by species $A$, $B$ and $C$ appears in red, green, and blue, respectively. In the color-coding of Fig.~\ref{fig1}(d), 
different levels of yellow, cyan and magenta correspond respectively to  a finite  fixation probability of  $A$ and $B$, $B$ and $C$,  $C$ and $A$, while white encodes the same fixation probability for each species ($\phi_{\rm i}\approx 1/3$). 

In Fig.~\ref{fig4}(a) we show results where the first reaction has power-law WTD~(\ref{eq:Psi-PL}) and the other reactions have exponential WTDs, with $\Lambda_{\rm A}$ given by~\eqref{lamPLmean}, and $\lambda_{\rm A}\!=\!Nk_{\rm A}a b$, $\lambda_{\rm B}\!=\!Nk_{\rm B}b c$ and $\lambda_{\rm C}\!=\!Nk_{\rm C}a c$. In Fig.~\ref{fig4}(b,c) we report results obtained for the dynamics where all reactions are drawn from a power-law WTD, with $\alpha_{\rm B}=\alpha_{\rm C}=10$ in (b), and $\alpha_{\rm B}=\alpha_{\rm C}=1.5$ in (c). Here, $\Lambda_{\rm B} = \lambda_{\rm B}/(\alpha_{\rm B} -1)$ and $\Lambda_{\rm C} = \lambda_{\rm C}/(\alpha_{\rm C} -1)$, see Eq.~(\ref{lamPLmean}). 
As a reference, it is useful to consider the fixation heatmap predicted by LOW for Markovian dynamics with exponential WTD:  when $k_{\rm B}=k_{\rm C}=1$, 
$A$ dominates (red phase) for $k_{\rm A}<1$ and $B$ and $C$ dominate (cyan phase) for $k_{\rm A}>1$, separated by $k_{\rm A}=1$ [dotted lines in Fig.~\ref{fig4}]. 

The  heatmap diagram of Fig.~\ref{fig4}(a) is mostly characterized  by a red phase dominated by $A$ ($\phi_{\rm A}\approx 1, \phi_{\rm B}\approx\phi_{\rm C}\approx 0$), and a cyan phase where $B$ and $C$ are the prevailing species ($\phi_{\rm A}\approx 0, \phi_{\rm B}\approx\phi_{\rm C}\approx 1/2$). The border between these phases is  an increasing function of $\alpha_{\rm A}$. When  $\alpha_{\rm A}$ approaches 1, the dynamics of \eqref{REPL} are not necessarily  characterized by closed orbits, 
and a third phase, not predicted by the LOW, emerges in blue: it corresponds to 
the 
dominance of $C$ ($\phi_{\rm A}\approx \phi_B  \approx 0,\phi_C\approx 1$), with  breaking of the $B/C$ symmetry. Here, as $\alpha_{\rm A}\to 1$, the typical interevent time (median) diverges [see Eq.~(\ref{eq:mean-pl})]. As a result, the typical production rate of $A$ individuals is very high at the expense of $B$ individuals. Thus, the population of $C$
can grow almost  without opposition from  its predator, species $B$, that is rapidly consumed by $A$, and hence $C$ eventually fixate the entire population when $\alpha_{\rm A}\approx 1$. Notably, Eqs.~(\ref{REPL}) support this analysis: as $\alpha_{\rm A}\to 1$,  memory kernel (\ref{MKPL}) becomes very large, $\Theta_{\rm PL}\gg 1$, which yields $c^*\to 1$, and  $a^*,b^*\to 0$. 
In contrast, when $\alpha_{\rm A}\gg 1$, we recover the LOW predictions and the separation between the red and cyan phases occurs around $k^*$ (dashed line) given by  \eqref{kstar-PL}. Remarkably, the prediction of $k^*$ as a separating curve turns out to be valid also at $\alpha_{\rm A}\gtrsim 1$. Here, as $\alpha_{\rm A}$ approaches $1$ and the median increasingly deviates from the mean, a striking departure from the LOW is observed; i.e., 
it is necessarily to significantly lower $k_{\rm A}$ (much below 1) for $A$ to win.
The diagram of Fig.~\ref{fig4}(b) is quantitatively similar to that of  Fig.~\ref{fig4}(a). This is because for large values of $\alpha_{\rm B},\alpha_{\rm C}$,  the power-law WTDs for the corresponding reactions are close to 
the exponential WTDs considered in Fig.~\ref{fig4}(a).

The heatmap of  Fig.~\ref{fig4}(c)
is characterized by the same phases as in Fig.~\ref{fig4}(a,b), with some major quantitative differences. In particular, we notice that  
the  separation between the cyan and red phases in panel (c) occurs for values of $k_{\rm A}$ much higher than 1 (predicted by the LOW). This stems from the typical rates of the last two predator-prey  reactions of \eqref{eq:reac} being less than their corresponding means, 
giving rise to the fixation of $A$  even 
for $k_{\rm A}\!>\!1$.

\vspace{0.2cm}\noindent\textbf{Fixation heatmaps for the gamma WTD.}
In Fig.~\ref{fig4}\black{(d-f)} we report the fixation heatmaps for the zRPS dynamics with 
gamma WTD for the reaction $A+B\to A+A$ as a function of $k_{\rm A}$ and $\alpha_{\rm A}$. Here, the WTD is given by \eqref{eq:Psi-G} and the parameters $(\Lambda_{\rm A}, \alpha)$ satisfy Eq.~(\ref{lamgammamean}). In panel (d), the WTDs of the other two reactions are exponential. 
As expected, the survival/fixation behavior
reproduces  the LOW scenario at  $\alpha_{\rm A} = 1$: with an  
$A$-dominated (red) phase where $k_{\rm A}<1$ and a (cyan) phase dominated by $B/C$  where $k_{\rm A}>1$.  In  Fig.~\ref{fig4}(d,e) the 
 white dotted line separates the 
two phases predicted by the LOW.
This has to be contrasted with the critical value (\ref{kstar-full}),
shown as the dashed white curve, which 
separates the phases where $A$ (red) dominates and where it does not dominate (blue/cyan region). In  Fig.~\ref{fig4}(d,e), the non-Markovian dynamics results in the fixation of $A$ where $k_{\rm A}(\alpha_{\rm A}) < k_{\rm A}^*(\alpha_{\rm A})$ which is the region of the phase space where
$c^* < 1/3$, see Eq.~(\ref{coex-gamma-full}). In this red region of the parameter space, the coexistence equilibrium is thus closest to the A-B edge of Fig.~\ref{fig1}, and species $A$ is hence the most likely to fixate the population. 
The opposite occurs when $k_{\rm A}(\alpha_{\rm A}) > k_{\rm A}^*(\alpha)$: species $B$ or $C$ prevail and $A$ goes extinct. 
While the zRPS with gamma WTD reproduces the LOW predictions for $\alpha_{\rm A}$ close to $1$, with $B$ and $C$ most likely to prevail with the same probability where $k_{\rm A}> 1$, the  $B/C$ symmetry is broken when  $\alpha_{\rm A}$ is distinctly below $1$ and in this case  $C$ is the most likely to prevail as indicated by the blue phase in Fig.~\ref{fig4}(d-f).

Notably, the striking  
symmetry-breaking effect in the case of the gamma WTD is not predicted by the LOW and is not captured by the MF approximation. We note that a similar, but less striking, effect is  
also  observed with  power-law WTD in the narrow region where $\alpha_{\rm A}\to 1$, see Fig.~\ref{fig4}(a-c). 
We conjecture that the stark contrast between the power-law and gamma WTD results stem from the ratio of the median to the mean of the WTD. Indeed, for the gamma WTD as $\alpha_{\rm A}$ decreases below $1$, the ratio between the median and mean goes to zero much more rapidly than in the power-law case (as $\alpha_{\rm A}$ goes to $1$), see e.g., Fig.~\ref{fig2}. Hence, the extreme scenario of almost complete depletion of  $B$ and takeover by $C$ species occurs much earlier with the gamma WTD.

We notice that  the nontrivial curve of $k_{\rm A}=k^*$ given by \eqref{kstar-full} determines  the separation between the phases where species $A$ dominates (red) and where it loses (blue/cyan) 
with excellent accuracy. While for practical reasons, the numerical simulations are limited to $\alpha_{\rm A}\leq 1$ (see Appendix C), we expect that the phases in which $A$ is dominant and where it loses is determined by $k^*$ also for $\alpha_{\rm A}>1$. 
In fact, when  $\alpha_{\rm A}>1$, the gamma WTD is unimodal, with the mean and median being increasingly closer as $\alpha_{\rm A}$ increases, and coinciding when $\alpha_{\rm A}\to \infty$. Thus, when $\alpha_{\rm A}>1$,  $A$ can prevail also for $k_A>1$, assuming that the fixation dynamics are qualitatively similar to those in the $\alpha\lesssim 1$ regime. In particular, at $\alpha_{\rm A}\to \infty$, $A$  prevails as long as $k<2/\ln 3\simeq 1.82$, see Eq.~(\ref{coex-gamma-full}).

In  Fig.~\ref{fig4}\black{(e,f)}, the last two reactions of \eqref{eq:reac}  occur with interevent times that are also distributed according to a gamma WTD [Eq.~(\ref{eq:Psi-G})] with the equivalent of \eqref{lamgammamean} for $\Lambda_{\rm B}$ and $\Lambda_{\rm C}$. In panel (e) we take $\alpha_{\rm B}=\alpha_{\rm C}= 0.9$ and the resulting heatmap is very similar to that in (d). Yet, in panel \black{(f)}  $\alpha_{\rm B}=\alpha_{\rm C}= 0.5$ and the separating interface markedly change.
Even when $k_{\rm A}=k_{\rm B}=k_{\rm C}=1$,  the values of $ \alpha_{\rm B}$  and $\alpha_{\rm C}$, hence the shape of the WTDs, change the range of $k_A$ for which the 
LOW predictions are  reproduced. In particular, when $\alpha_{\rm B}=\alpha_{\rm C} = 0.5$, species $A$ dominates  for higher values of $k_{\rm A}$ than under Markovian dynamics, and thus, the red region in panel \black{(f)} is larger than in (d,e) [see also (a)-(c)]. We also notice that for $k_{\rm A}\gg 1$ \black{and $\alpha_{\rm A} > \alpha_{\rm B}, \alpha_{\rm C}$} a green $B$-dominated phase appears in  Fig.~\ref{fig4}\black{(f)}. \black{Notably, both cases of symmetry breaking---dominance of $C$ for $\alpha_{\rm A} < \alpha_{\rm B}, \alpha_{\rm C}$, and $B$ for $\alpha_{\rm A} > \alpha_{\rm B}, \alpha_{\rm C}$---are not captured by MF theory. Thus, the observed symmetry breaking and dominance of either of the species may strongly depend on the full set of WTDs of all reactions rather than on a simple hierarchy.}

\begin{figure}[t!]
\centering
\includegraphics[width=.95\linewidth]{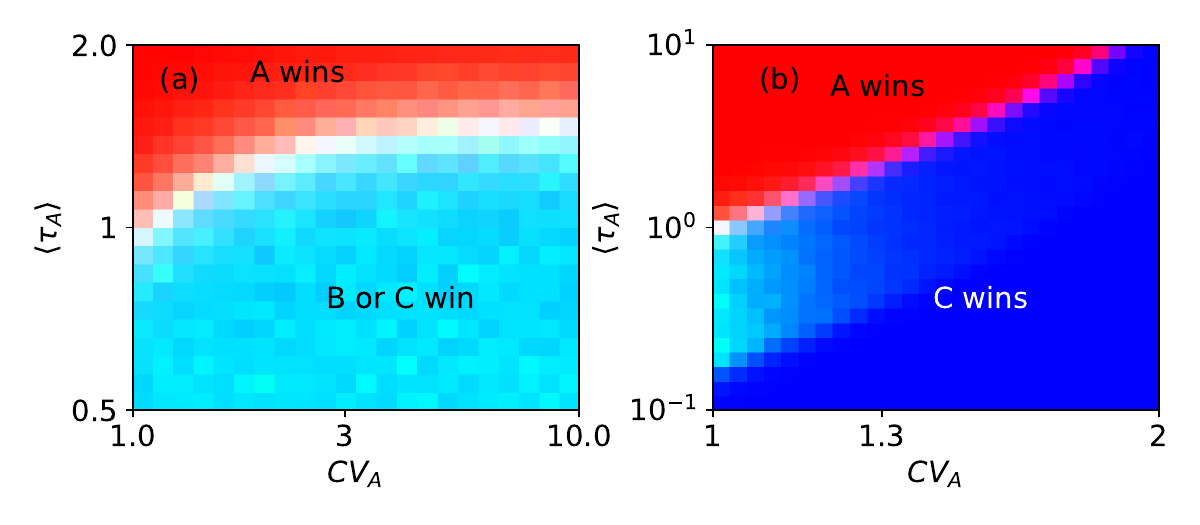}
\vspace{-5mm}
\caption{RGB fixation heatmaps \black{(based on the diagram of Fig.~\ref{fig1})} for  power-law (a) and gamma (b) WTDs
of the first reaction of \eqref{eq:reac}: mean interevent time  $\langle \tau_{\rm A}\rangle$ vs.  coefficient of variation ${\rm CV_{\rm A}}$.
Here $N \!=\! 999$, $k_{\rm B} \!=\! k_{\rm C} \!=\! 1$, and $\alpha_{\rm A}\!\leq \!1$ in (b). }
\label{fig5}
\end{figure}

\vspace{0.1cm}\noindent\textbf{Comparison of power-law and gamma WTDs.}
To further compare the effect of the power-law and gamma WTD on the RPS survival scenarios,
we plot in Fig.~\ref{fig5} the fixation maps under power-law and gamma WTDs versus the average waiting time $\langle \tau_{\rm A}\rangle$
and  coefficient of variation ${\rm CV}_{\rm A}$. This allows us to directly compare the effect of these different  WTDs. 
As expected, in both panels for ${\rm CV}_{\rm A}= 1$ we fully reproduce the predictions of the LOW for exponential WTDs:  species $A$ is the most likely to fixate the population (red phase) when $\langle \tau_{\rm A}\rangle>1$ ($k_{\rm A}<1$), whereas  species $B$ and $C$ are the most likely to survive (same probability) and $A$ goes extinct (cyan phase), when $\langle \tau_{\rm A}\rangle<1$ ($k_{\rm A}>1$).
When ${\rm CV}_{\rm A} > 1$, the survival scenarios drastically deviate  from the LOW predictions for the two cases considered here.
When the first reaction has a power-law WTD, the whitish interface (equal fixation probability for all species)  is a concave function which gradually changes as ${\rm CV}_{\rm A}$ grows,  see Fig.~\ref{fig5}(a). Here, as ${\rm CV}_{\rm A}$ is increased,  larger $\langle \tau_{\rm A}\rangle$ (smaller  $k_{\rm A}$) is required for  $A$ to win, with saturation of $\langle \tau_{\rm A}\rangle$ when ${\rm CV}_{\rm A}\gg 1$. A much more pronounced effect is observed in the case of the gamma WTD with $\alpha_{\rm A}\leq 1$ shown in Fig.~\ref{fig5}(b). Here, the whitish interface is a convex function and is much steeper than in the power-law case, with no observable saturation. Remarkably, when ${\rm CV}_{\rm A}$ grows by a factor of $2$, in order for $A$ to still win, $\langle \tau_{\rm A}\rangle$ ($k_{\rm A}$) needs to increase (decrease) by a factor of $> 10$,  see Fig.~\ref{fig5}(b). A similar increase in  $\text{CV}_{\rm A}$ in the power-law case leads to increase in $\langle \tau_{\rm A}\rangle$  of only 
$\sim 40\%$,  see Fig.~\ref{fig5}(a). 

Moreover, as also observed in Fig.~\ref{fig4}, in the power-law case the whitish interface separates fixation of $A$ (red regime) and extinction of $A$ accompanied by equal fixation probability of $B$ and $C$ (cyan regime). In contrast, in the case of gamma WTD with $\alpha_{\rm A}\leq 1$, the interface separates fixation of $A$ (red regime) and $C$ (blue regime), and symmetry between $B$ and $C$ is broken. This is because for the gamma WTD, increasing  ${\rm CV}_{\rm A}$ has a much stronger effect on decreasing the typical interevent time than in the power-law case, see  Fig.~\ref{fig5}.

\vspace{0.4cm}\noindent\textbf{\Large{Discussion}}\\
\noindent
There is mounting evidence  that  waiting times between  successive events play an important role in shaping evolutionary processes across biology and ecology. For instance, optimal foraging strategies have been linked to heavy-tailed waiting time distributions (WTDs)~\cite{Viswanathan99,Metzler14,Wosniack17,Guinard21,vilk2022ergodicity}. Moreover, with the development of microfluidic devices and single-cell experiments, the role of the reproduction time distribution on microbial growth has received significant attention \cite{Amir19,Jafarpour23,Nozoe20}.  
Here, we studied the influence of WTDs on the fate of the paradigmatic zero-sum rock-paper-scissors (zRPS) game between three species in cyclic competition, which is broadly used in biology and ecology
~\cite{Sinervo_96,Frach96,Hofbauer,Frach98,Nowak,Reich2006,Reich2007,Perc_2007,Szabo_2007,Jiang_2011,Broom,Review2014,Szolnoki2016,Menezes2018,Review2018,West2018,West2020,Liao2020,Kerr2002}. The zRPS dynamics are classically
modeled in terms of Markov processes, with exponential WTDs, and its final state obeys the simple ``law of the weakest'' (LOW)~\cite{Berr2009} stating that, the species that is most likely to  fixate  is the one with lowest  predation-reproduction (predator-prey) rate. 
Here we have shown that the LOW predictions are drastically altered in the non-Markovian zRPS model with non-exponential WTDs.

By combining analytical arguments and 
extensive stochastic simulations, we investigated the 
fixation probability of each species when at least one of the zRPS reactions has a non-exponential WTD, and we have focused  on the two-parameter power-law and gamma WTDs.  The former is related to anomalous diffusion~\cite{Metzler14}, abundantly found in animal behavior, while the latter is often used to model the reproduction of microbial cells~\cite{Amir19,Jafarpour23}. 
Keeping the same mean for all WTDs, we found that the fate of the zRPS dynamics is drastically affected by the features of the non-exponential WTD:
the conditions under which one species is most likely to fixate depend non-trivially on the WTD parameters in addition to  the reaction rates.  We visualized our findings in  heatmap fixation diagrams identifying the parameter regions dominated by each species. Depending on the WTD parameters, the phase
in which one species dominates over the others can be enhanced or reduced with respect to the predictions of the LOW, see Figs.~\ref{fig4} and \ref{fig5}. 

The major deviations from the LOW arise when the difference between the ``typical'' and ``mean'' interevent times (difference between the median and  mean of the WTD) increases. By focusing on positively skewed distributions (like the exponential WTD), we showed that the region of dominance of the species whose reproduction is governed by the non-exponential WTD strongly depends on the WTD coefficient of variation (CV). The region of dominance of each species thus shrinks or grows with respect to the LOW predictions, depending on whether ${\rm CV}\!>\!1$ or ${\rm CV}\!<\!1$; e.g., the phase dominated by species $A$ shrinks as ${\rm CV}\!>\!1$, see Fig.~\ref{fig5}. 
In addition, the symmetry between the other two species, a signature of exponential WTDs, is expected to be broken when the ratio between the median and mean vanishes. 
Our analytical arguments are based on the analysis of  generalized MF rate equations. These involve   memory kernels derived from the underlying non-Markovian master equations.  While it would be desirable to establish a physical interpretation of the memory kernels, this is beyond the scope of this study. Yet, 
our work is readily applicable for a wide class of non-exponential WTDs, \black{including scenarios in which species switch between multiple internal behavioral states, each having a distinct WTD,} or empirical distributions which are directly inferred from data.

We believe that our analysis of 
the influence of non-exponential WTDs on the fate of the zRPS model 
can help shed further light on non-Markovian evolutionary dynamics, and can help motivate  further studies. Besides 
a physical derivation of the memory kernels, understanding  how  the mean fixation time of each species  is affected by general non-exponential WTDs is an open question. Moreover,  as a number of  microbial experiments  have been  modeled using RPS dynamics~\cite{Kerr2002,Liao2020},  we expect that the effects of interevent time distributions on species in cyclic competition 
can be tested  in laboratory-controlled experiments by employing our theoretical approach.

\vspace{0.2cm}
\noindent{\bf Data availability statement:}
 Data and codes  that support our work are  available at the following \href{https://gitlab.com/ohad.vilk/non_exponential_rock_paper_scissors}{URL}~\cite{gitlab-data}.

\vspace{0.2cm}
\noindent{\bf Acknowledgments:}
M.A. and O.V. acknowledge support from ISF grant 531/20. 
M.M. thankfully acknowledges  support of the EPSRC Grant No.
EP/V014439/1.
\vspace{.5cm}

\bibliography{biblio}

\subsection*{Appendix A: Stability analysis in the non-Markovian case}
\label{stability}
\setcounter{equation}{0}
\renewcommand{\theequation}{A\arabic{equation}}

Here we briefly study the linear stability of the coexistence equilibrium $(a^*,b^*,c^*)$ in the case of non-exponential WTD for the first reaction.
As stated in the main text, with exponential WTDs, the MF rate equations (\ref{eq:MF})
admit  the constant of motion (\ref{eq:CoM}). In the phase space, the MF dynamics  are therefore characterized by closed orbits surrounding \eqref{eq:sstar} that 
is a (nonlinear) center, see Fig.~\ref{fig1}(a). In fact, the Jacobian of  \eqref{eq:MF} evaluated at \eqref{eq:sstar} has two conjugate purely imaginary eigenvalues, denoted by $\{\beta i,-\beta i\}$ where $\beta$ depends on the $k_i$ values. In the  case of $k_{\rm A}=k_{\rm B}=k_{\rm C}=1$,  $\beta=-1/\sqrt{3}$. 

Interestingly, when the first reaction of \eqref{eq:reac} has a power-law WTD and $\alpha_{\rm A}\gg 1$, the Jacobian of   \eqref{REPL} evaluated at 
(\ref{coex-pl_full}) also has a pair of conjugate purely imaginary eigenvalues of the form 
 $\{(\beta+\beta_{\text{PL}}/\alpha_{\rm A})i,-(\beta+\beta_{\text{PL}}/\alpha_{\rm A})i\}$, where $\beta_{\text{PL}}$
depends on $k_i$.  In the  case of $k_{\rm A}=k_{\rm B}=k_{\rm C}=1$, $\beta$ is identical to the exponential case, and $\beta_{\rm PL}=-1/(6\sqrt{3})$. Moreover,
the quantity (\ref{eq:CoM}) 
is conserved by \eqref{REPL} 
to leading order in $1/\alpha_{\rm A}$: $d{\cal R}/dt={\cal R}(\Theta_{\rm PL} -1)k_{\rm A}(k_{\rm B}b-k_{\rm C}a)={\cal O}({\cal R}/\alpha_{\rm A})$, where we have used \eqref{eq:thetaPLpprox}.
This indicates that when $\alpha_{\rm A}\gg 1$, the phase space dynamics prescribed by \eqref{REPL} 
are characterized by closed orbits surrounding the equilibrium  
(\ref{coex-pl_full}), where $k_{\rm B}b^*-k_{\rm C}a^*=0$, for long transients, see Fig.~\ref{fig1}(b). This behavior is qualitatively similar to that predicted by  \eqref{eq:MF}. However, the location of $(a^*,b^*,c^*)$ and the shape of the orbits around it now depend on the non-exponential WTD parameter $\alpha_{\rm A}$, yielding deviations from the  
survival / fixation scenarios predicted by the LOW, see  main text.
While this analysis cannot be extended for arbitrary values of $\alpha_{\rm A}$, our extensive numerical simulations have confirmed that $(a^*,b^*,c^*)$ is a nonlinear center for most values $\alpha_{\rm A}>1$, see Fig.~\ref{fig1}(b).

A similar analysis can be done in the case of gamma WTD, when $\alpha_{\rm A}$ is close to $1$. Introducing $\epsilon=\alpha_{\rm A}-1$, for $|\epsilon|\ll 1$, the coexistence equilibrium 
$(a^*,b^*,c^*)$, given by \eqref{coex-gamma-full}, is again a center associated with two purely imaginary conjugate eigenvalues,  
$\{(\beta+\beta_{\text{G}}\epsilon) i,-(\beta+\beta_{\text{G}}\epsilon) i\}$. For $k_{\rm A}\!=\!k_{\rm B}\!=\!k_{\rm C}\!=\!1$, $\beta$ is identical to the exponential case, and $\beta_{\rm G}=(3\ln 3\!-\!2)/(6\sqrt{3})$.  
In fact, the 
quantity (\ref{eq:CoM}) 
is again conserved by the generalized rate equations \eqref{REPL} (with \eqref{memory_gamma})
to leading order in $\epsilon$ when $\alpha \approx 1$: $d{\cal R}/dt={\cal R}(\Theta_{\rm G} -1)k_{\rm A}(k_{\rm B}b-k_{\rm C}a)={\cal O}({\cal R}\epsilon)$, where we have used \eqref{thetagamapprox}.
This again indicates that when $|\epsilon|\ll 1$, the phase space dynamics prescribed by \eqref{REPL} 
are characterized by closed orbits surrounding  
(\ref{coex-gamma-approx}), see Fig.~\ref{fig1}(c). 
Thus, the 
survival scenario can again be inferred from the location of $(a^*,b^*,c^*)$. 

Notably, in the regime of $\alpha_{\rm A}={\cal O}(1)$ in the power-law case, and $|\alpha_{\rm A}-1|={\cal O}(1)$ in the gamma case, we cannot prove in general that the dynamics include closed orbits. Nevertheless, our extensive numerical simulations show that, as long as $\alpha$ is not to close to $1$ (in the power-law case) and to $0$ (in the gamma case), closed orbits around the equilibrium state are still observed.

\subsection*{Appendix B: Generalized rate equations}
\label{AppC}
\setcounter{equation}{0}
\renewcommand{\theequation}{B\arabic{equation}}

In this appendix, we outline the derivation of the generalized rate equations (\ref{REPL}) with the memory kernel (\ref{MKPL}) in the case of a power-law WTD~(\ref{eq:Psi-PL}), and with the memory kernel~(\ref{memory_gamma}) in the case of a gamma WTD~(\ref{eq:Psi-G}).
In the following, we respectively number the reactions $A+B\to A+B$, $B+C\to B+B$ and $C+A\to C+C$ as the first, second and third reactions, and henceforth denote their WTDs by $\psi_{\rm A}(\tau_{\rm A})$, $\psi_{\rm B}(\tau_{\rm B})$ and $\psi_{\rm C}(\tau_{\rm C})$.

We begin by writing the master equation for the probability $P_{n_{\rm A},n_{\rm B},n_{\rm C}}(t)$ to find $n_{\rm A}$, $n_{\rm B}$ and $n_{\rm C}$ individuals of type $A$, $B$, and $C$, at time $t$. For Markovian dynamics, where all reactions have  exponential WTDs, one obtains:\begin{eqnarray}\label{MEexp}
    &&\hspace{-3mm}\dot{P}_{\textbf{n}}(t)=\frac{k_{\rm A}}{N}\left(E_{n_{\rm A},n_{\rm B}}^{-1,+1}-1\right)n_{\rm A}n_{\rm B}P_{\textbf{n}}(t)\\
    &&\hspace{-3mm}+\frac{k_{\rm B}}{N}\!\left(E_{n_{\rm B},n_{\rm C}}^{-1,+1}\!\!-\!1\right)\!n_{\rm B}n_{\rm C}P_{\textbf{n}}(t)\!+\!\frac{k_{\rm C}}{N}\!\left(E_{n_{\rm A},n_{\rm C}}^{+1,-1}\!\!-\!1\right)\!n_{\rm A}n_{\rm C}P_{\textbf{n}}(t),\nonumber
\end{eqnarray}
where $N=n_{\rm A}+n_{\rm B}+n_{\rm C}$ is the total (constant) population size. Here, we have defined a step operator for brevity of notation, $E_{k_1,k_2}^{j_1,j_2} f(k_1,k_2) = f(k_1 + j_1,k_2+j_2)$, and denoted the population sizes by a vector $\textbf{n}=\{n_{\rm A},n_{\rm B},n_{\rm C}\}$. In the general case of reactions with non-exponential WTDs, the master equation becomes non-local in time; i.e., the current state depends on the entire history of the process with prescribed memory kernels that depend on the WTDs of the different processes, see below. In this case, Eq.~(\ref{MEexp}) becomes
\begin{eqnarray}\label{MEnonexp}
    &\dot{P}_{\textbf{n}}(t)=k_{\rm A}\left(E_{n_{\rm A},n_{\rm B}}^{-1,+1}-1\right)\!\int_0^t \!M_{\rm A}(\textbf{n},t')P_{\textbf{n}}(t-t')dt'\nonumber\\
    &+k_{\rm B}\left(E_{n_{\rm B},n_{\rm C}}^{-1,+1}-1\right)\!\int_0^t \!M_{\rm B}(\textbf{n},t')P_{\textbf{n}}(t-t')dt'\nonumber\\
    &+k_{\rm C}\left(E_{n_{\rm A},n_{\rm C}}^{+1,-1}-1\right)\!\int_0^t \!M_{\rm C}(\textbf{n},t')P_{\textbf{n}}(t-t')dt'.
\end{eqnarray}
Here $M_{\rm A}(\textbf{n},t)$, $M_{\rm B}(\textbf{n},t)$ and $M_{\rm C}(\textbf{n},t)$ are the memory kernels for the creation of $A$, $B$ and $C$, respectively, and the constant $N$ was absorbed in these kernels. Following the derivation in \cite{aquino2017chemical,vilk2024non,vilk2024escape} we find the memory kernels by Laplace-transforming Eq.~(\ref{MEnonexp}). We first define the probability density for the first reaction to occur at time $t$ while the other two reactions \textit{do not} occur until $t$:
\begin{eqnarray}\label{phis}
    \Phi_{\rm A}(t) = \psi_{\rm A}(t) \int_{t}^{\infty}\psi_{\rm B}(\tau)d\tau\int_{t}^{\infty}\psi_{\rm C}(\tau)d\tau,
\end{eqnarray}
where $\Phi_{\rm B}(t)$ and $\Phi_{\rm C}(t)$ are defined similarly.
Note that, in addition to their time dependence,  $\Phi_{\rm A}$, $\Phi_{\rm B}$ and $\Phi_{\rm C}$ may also depend  on $n_{\rm A}$, $n_{\rm B}$ and $n_{\rm C}$.
 It can be shown that the memory kernels in Laplace space satisfy~\cite{aquino2017chemical,vilk2024non,vilk2024escape}:
\begin{equation} \label{Mkernel_FullModel}
   \hspace{-3mm}\tilde{M}_{\rm X}(s)  = s \tilde{\Phi}_{\rm X}(s)\!\left[1 - \tilde{\Phi}_{\rm A}(s)-\tilde{\Phi}_{\rm B}(s)-\tilde{\Phi}_{\rm C}(s)\right]^{-1}\!\!\!,
\end{equation}
where $X=\{A,B,C\}$ and $\tilde{\Phi}$ denotes the Laplace transform of Eqs.~(\ref{phis}), and $s$ is the Laplace variable.

We now explicitly compute the memory kernels in the case of power-law WTD for the first reaction given by Eq.~(\ref{eq:Psi-PL}) with $\Lambda_{\rm A}=\lambda_{\rm A}/(\alpha_{\rm A}-1)$, and exponential WTDs for the second and third reactions, such that $\psi_{\rm B}(\tau)=\lambda_{\rm B} e^{-\lambda_{\rm B}\tau}$ and $\psi_{\rm C}(\tau)=\lambda_{\rm C} e^{-\lambda_{\rm C}\tau}$. Computing  the Laplace-transforms of Eqs.~(\ref{phis}), $\tilde{\Phi}$,  plugging the result into Eqs.~(\ref{Mkernel_FullModel}), putting $\lambda_{\rm A}=k_{\rm A} n_{\rm A} n_{\rm B}/N$, $\lambda_{\rm B}=k_{\rm B} n_{\rm B} n_{\rm C}/N$ and $\lambda_{\rm C}=k_{\rm C} n_{\rm A} n_{\rm C}/N$, and taking the leading-order result with respect to $N\gg 1$, one obtains $\tilde{M}_{\rm B}(s)= N k_{\rm B}b \,c$ and $\tilde{M}_{\rm C}(s)= N k_{\rm C}a\, c$, where we have used the fractions $a=n_{\rm A}/N$, $b=n_{\rm B}/N$ and $c=n_{\rm C}/N$. In addition, we find $\tilde{M}_{\rm A}(s)= Nk_{\rm A}a\,b \,\Theta_{\text{PL}}(a,b,c)+{\cal O}(s)$, where $\Theta_{\text{PL}}(a,b,c)$ is given by Eq.~(\ref{MKPL}). As all the memory kernels are constant in $s$ in the leading order, performing an inverse Laplace-transform yields to leading order: 
\begin{eqnarray}
    &&M_{\rm A}(a,b,c,t)=Nk_{\rm A}a\,b \,\Theta_{\text{PL}}(a,b,c)\,\delta(t),\\
    &&M_{\rm B}(a,b,c,t)=Nk_{\rm B}b\,c \,\delta(t),\;\;M_{\rm C}(a,b,c,t)=Nk_{\rm C}a\,c \,\delta(t).\nonumber
\end{eqnarray}
Plugging the three memory kernels into the master equation [Eq.~(\ref{MEnonexp})], all the integrals over time yield the integrands evaluated at time $t$, and one obtains:
\begin{eqnarray}\label{MEeff}
   & \hspace{-2cm}\dot{P}_{\textbf{n}}(t)=\frac{k_{\rm A}}{N}\left(E_{n_{\rm A},n_{\rm B}}^{-1,+1}\!-\!1\right)n_{\rm A}n_{\rm B}\Theta_{\text{PL}}(\textbf{n})P_{\textbf{n}}(t)\\
    &+\frac{k_{\rm B}}{N}\!\left(E_{n_{\rm B},n_{\rm C}}^{-1,+1}\!\!-\!1\right)\!n_{\rm B}n_{\rm C}P_{\textbf{n}}(t)+\frac{k_{\rm C}}{N}\!\left(E_{n_{\rm A},n_{\rm C}}^{+1,-1}\!\!-\!1\right)\!n_{\rm A}n_{\rm C}P_{\textbf{n}}(t)\!.\nonumber
\end{eqnarray}
This equation coincides with master equation~(\ref{MEexp}) up to the factor of $\Theta_{\text{PL}}(\textbf{n})$, which is the signature of the non-Markovian nature of the first reaction.
As a result, and using the definition of the species averages 
\begin{equation}
    \hspace{-3mm}\bar{n}_{\rm A}\!=\!\!\sum_\textbf{n}\!n_{\rm A}P_{\textbf{n}}(t),\;\;\bar{n}_{\rm B}\!=\!\!\sum_\textbf{n}\!n_{\rm B}P_{\textbf{n}}(t),\;\;\bar{n}_{\rm C}\!=\!\!\sum_\textbf{n}\!n_{\rm C}P_{\textbf{n}}(t),
\end{equation}
in Eq.~(\ref{MEeff}), we arrive at rate equations~(\ref{REPL}) for $\bar{a}=\bar{n}_{\rm A}/N$, $\bar{b}=\bar{n}_{\rm B}/N$ and $\bar{c}=\bar{n}_{\rm C}/N$,  with $\Theta_{\text{PL}}(a,b,c)$ given by Eq.~(\ref{MKPL}). Note that, in the MF limit, $N\to \infty$, the average species fractions  $\bar{a}, \bar{b}, \bar{c}$ coincide with the fractions 
$a=n_{\rm A}/N$, $b=n_{\rm B}/N$ and $c=n_{\rm C}/N$, and, for brevity, the latter have been used in all the MF equations in the main text [e.g., Eqs.~(\ref{eq:MF}) and (\ref{REPL})].

Similarly, the memory kernel and rate equations for the gamma WTD can be found via Eq.~(\ref{eq:Psi-G}) for the first reaction with $\Lambda_{\rm A}\!=\!\lambda_{\rm A}\alpha_{\rm A}$, and  repeating the above steps.


\subsection*{Appendix C: Computational methods}
\label{AppD}
\setcounter{equation}{0}
\renewcommand{\theequation}{C\arabic{equation}}

\subsection{Simulations}
Here we summarize the simulation methods we have used. We start with a description of the original Gillespie algorithm \cite{Gillespie76}, followed by the Laplace Gillespie algorithm \cite{masuda2018gillespie} used to simulate non-exponential WTD.  

\subsubsection{Gillespie Algorithm}

The original Gillespie algorithm assumes \(\mathcal{N}\) independent Poisson processes with rates \(\lambda_i\) (\(1 \leq i \leq \mathcal{N}\)) running in parallel. The combined effect of these Poisson processes results in a superposed Poisson process with a total rate \(\sum_{i=1}^\mathcal{N} \lambda_i\). The algorithm steps are as follows:

\begin{enumerate}
    \item 
\textit{Time Increment (\(\Delta t\)) Calculation}. 
The time to the next event in the superposed Poisson process, follows the exponential distribution:
\begin{eqnarray}
&\varphi(\Delta t) = \left( \sum_{i=1}^N \lambda_i \right) e^{-\left( \sum_{i=1}^N \lambda_i \right) \Delta t}.
\end{eqnarray}
Using the survival function, which is the probability that a random variable exceeds a given value:
\begin{eqnarray}
&\int_{\Delta t}^\infty \varphi(t') dt' = e^{-\left( \sum_{i=1}^N \lambda_i \right) \Delta t},
\end{eqnarray}
with $\Delta t \!=\!\! -\!\ln u\!\Big/\!\!\sum_{i=1}^N \!\!\lambda_i$
and \(u\!\in\![0,1]\)  uniformly chosen.
    \item 
\textit{Event Determination}. 
Identify process \(i\) that generated the event with probability: 
$\Pi_i = \lambda_i\Big/\!\!\sum_{i=1}^N \lambda_i.$
    \item 
\textit{Process Update}. 
Advance  time by \(\Delta t\) and repeat.
\end{enumerate}

\subsubsection{Laplace Gillespie Algorithm}

The Laplace Gillespie algorithm is designed for efficient simulation of non-Markovian point processes by utilizing an event-modulated Poisson process, see details in \cite{masuda2018gillespie}.  The key steps are as follows:

\begin{enumerate}
    \item 
Initialize each of the \({\cal N}\) processes by drawing the rate \(s_i\) (\(1 \leq i \leq {\cal N}\)) according to its density function \(p_i(s_i)\), defined in terms of the WTD $\psi(\tau)$: 
\begin{eqnarray}
   & \psi(\tau) = \int_0^\infty p(s) s e^{-s \tau} ds. 
\end{eqnarray}
Alternatively,  integrating both sides one can write 
\begin{eqnarray}
    &\Psi(\tau) = \int_\tau^\infty \psi(\tau')d\tau' = \int_0^\infty p(s) e^{-s \tau} ds. 
\end{eqnarray}
This entails that $p(s)$ is the inverse Laplace transform of the survival probability $\Psi(\tau)$.
    \item 
Draw the time until next event $\Delta t \!=\! -\!\ln u\!\Big/\!\!\sum_{j=1}^{\cal N} \!s_j$,
with  \(u\in[0,1]\)  uniformly chosen.
\item 
Select the process \(i\) that has generated the event with probability: $\Pi_i = s_i\Big/\!\!\sum_{j=1}^{\cal N} s_j$.
\item 
Draw a new rate \(s_i\) according to \(p_i(s_i)\). For any processes \(j\) (\(1 \leq j \leq {\cal N}\)) whose interevent time statistics have changed following the occurrence of the event in steps 2-3, update their rates \(\lambda_j\) according to modified \(p_j(s_j)\).
\item 
Repeat steps 2-4 or exit (e.g., upon fixation).

\end{enumerate}

For an exponential distribution~(\ref{eq:Psi-exp}), we have a Poisson process with rate \(s_0\); i.e., \(\psi(\tau) = s_0 e^{-s_0 \tau}\) is trivially generated by \(p(s) = \delta(s - s_0)\), where \(\delta\) is the delta function.
For power-law distribution following Eq.~(\ref{eq:Psi-PL}), \(p(\lambda)\) can be shown to follow a gamma distribution \cite{masuda2018gillespie} given by:
\begin{equation} \label{p_lam_for_PL}
p(s) = \frac{s^{\alpha-1} e^{-s/\Lambda_{\rm A}}}{\Gamma(\alpha) \Lambda_{\rm A}^{\alpha_{\rm A}}},
\end{equation}
where \(\Gamma(\alpha)\) is the gamma function, \(\alpha\) is the shape parameter, and \(\Lambda_{\rm A}\) is the scale parameter.
Similarly, for gamma distribution WTD with $\alpha < 1$, $p(s)$ is given by 
\begin{equation} \label{p_lam_for_gamma}
    \hspace{-2mm}p(s) = 
    \begin{cases}
        0 ,\;\; s < \Lambda_{\rm A} , \\
        \left[\Gamma(\alpha)\Gamma(1\!-\!\alpha)s( s/\Lambda_{\rm A}\!-\!1)^\alpha\right]^{-1}\!\!,\;\; s > \Lambda_{\rm A}.    
    \end{cases}
\end{equation}
Equations (\ref{p_lam_for_PL}) and (\ref{p_lam_for_gamma}) are used here to simulate the model with power-law and gamma WTD, respectively. 

Importantly, for the gamma WTD $p(s)$ can be used to express the inverse Laplace transform of $\Psi(\tau)$ only for $0< \!\alpha \!< 1$ \cite{masuda2018gillespie}. Thus, this algorithm cannot be used for gamma WTD with $\alpha > 1$, a regime that will be considered elsewhere with different computational techniques.  
 It is  worth noting that, since here ${\rm CV_{A}}(\tau)=\alpha_{\rm A}^{-1/2}$, it is precisely in the regime where $\alpha_{\rm A}<1$ where the WTD is wider than an exponential interevent distribution, which is the main focus of this study.

\begin{figure}[t]
\centering
 \includegraphics[width=.82\linewidth]{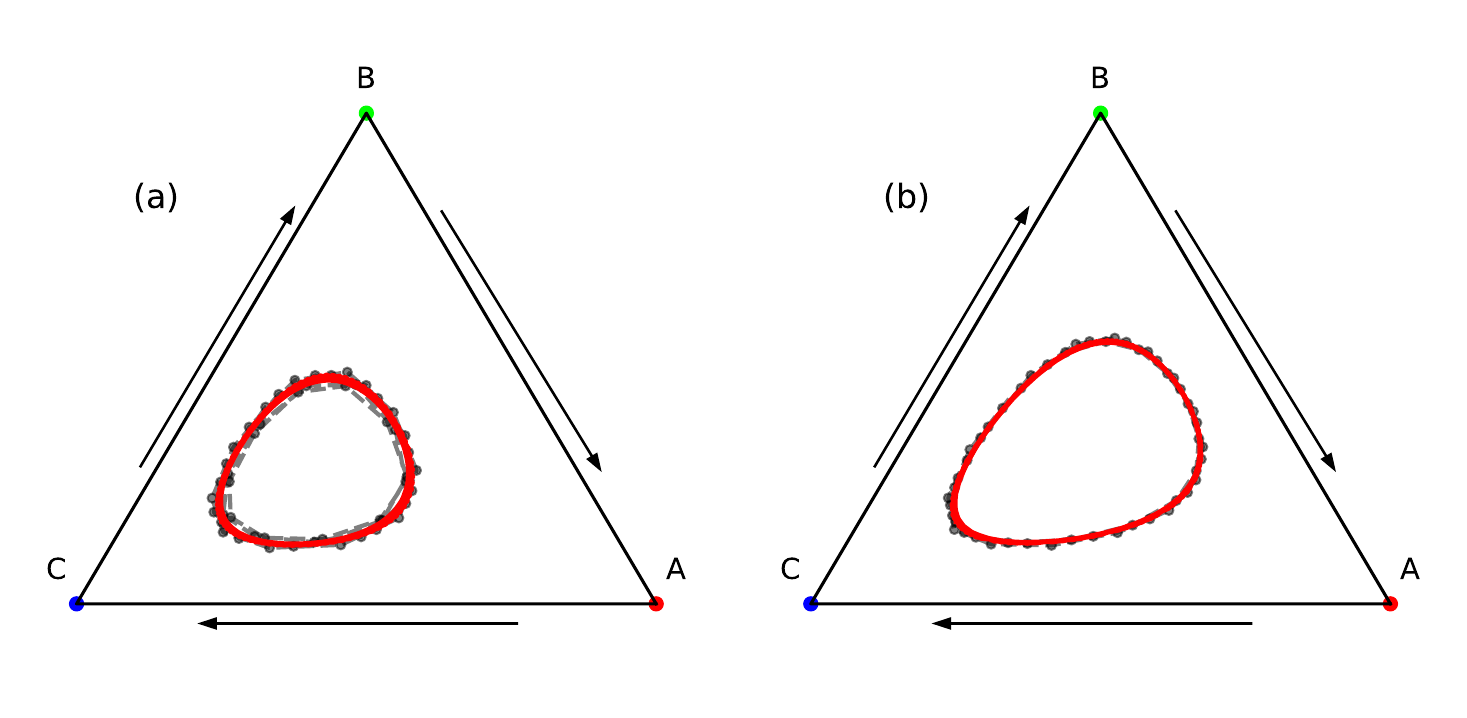}
\vspace{-6mm}
\caption{Dynamics in  
the ternary simplex for the model of zRPS with the last two reactions of  \eqref{eq:reac} having  exponential WTDs with $k_B=k_C=1$, while the
first reaction has 
a power-law WTD with $(k_A,\alpha_{\rm A})\!=\!(1,2)$ in (a),  and a gamma WTD with $(k_A,\alpha_{\rm A})\!=\!(0.4,\! 0.4)$ in (b). Gray dotted lines are stochastic  trajectories $(n_A,n_B,n_C)/N$, averaged over $10^4$ simulations starting from the same initial condition (clockwise dynamics), where trajectories are shown for $0<t\leq 100$ (omitting initial transients).  Red thick lines:
numerical solution of the deterministic rate equations. Corners  correspond to the fixation of the labeled species, where it has concentration 1. The initial concentrations are $1/3$ for each species.
}
\label{fig6}
\end{figure}

\begin{figure}[t]
\centering
 \includegraphics[width=1\linewidth,height=2.4cm]{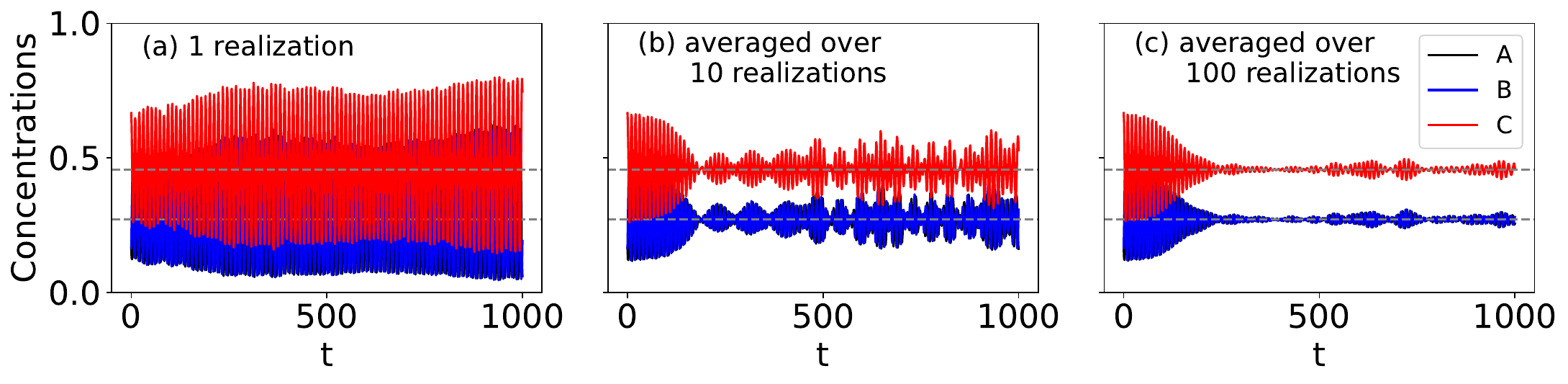}
\vspace{-7mm}\caption{Shown are concentrations of $A$ (blue), $B$ (black) and $C$ (red)  versus time
from stochastic simulations 
for the zRPS model \eqref{eq:reac} with the last two reactions of  \eqref{eq:reac} having exponential WTDs with $k_B=k_C=1$, while the first reaction has a power-law WTD with $(k_A,\alpha_{\rm A})\!=\!(1,2)$. The solid lines in (a) represent a single realization, while in (b) and (c) they represent ensemble averages over 10 and 100 simulations, respectively. The dashed lines are predictions of Eq.~(\ref{REPL}). In all panels the total population is $N = 3\cdot 10^4$ and the initial concentrations of $A$ and $B$ are $1/6$ while that of $C$ is $2/3$; also note that, the concentrations of $A$ and $B$ are overlapping. 
} 
\label{fig7}
\end{figure}

\subsection{Numerical methods}
Here we explain the analysis used to generate Figs.~\ref{fig1}, and \ref{fig3}. 
The deterministic orbits plotted in Fig.~\ref{fig1} (red solid lines) are obtained by numerically solving rate equations~(\ref{REPL}) for both power-law and gamma WTD. In the exponential case [Fig.~\ref{fig1}(a)] we plot the outermost orbit defined by ${\cal R}(t)=1/N$, where ${\cal R}$ is given by \eqref{eq:CoM}. For the non-exponential WTD [Fig.~\ref{fig1}(b-c)], ${\cal R}(t)$ is not necessarily constant and the outermost orbit may not admit a closed form expression. We show the differences between the exponential and non-exponential cases by  solving Eqs.~(\ref{REPL}) with an identical initial condition ${\cal R}(t = 0)=1/N$, which resides on the outermost orbit of the exponential case in \ref{fig1}(a). 
The conserved quantity in Fig.~\ref{fig1}(a),  ${\cal R}(t)=abc^{0.8}=1/100$, is no longer conserved as can be seen from the clear differences between the red solid lines in Figs.~\ref{fig1}(a) and \ref{fig1}(b-c). 

In Fig.~\ref{fig3}   we compare the theoretical fixed point to the fixed points of the stochastic simulations. To obtain the latter we performed $10^4$ simulation with $N = 10^5$, for times $t = 10^3 \ll N$, such that nearly no simulation reaches fixation. 
The steady-state concentrations of $A$, $B$ and $C$ in the stochastic dynamics are obtained by averaging each population over all data points in all simulations.
Notably, by averaging over all simulations at constant time intervals it is also possible to obtain the dynamic orbits from the stochastic simulations.  A typical example of the orbits surrounding the coexistence equilibrium $(a^*,b^*,c^*)$ in the ternary simplex  is reported in Fig.~\ref{fig6} for the power-law and gamma WTDs with typical parameters considered in this work. Here, all simulations have the same initial concentration of species. In this figure we find good agreement between the averaged stochastic trajectories and numerical  solutions of the deterministic rate equation for both classes of WTDs.  Finally, in Fig.~\ref{fig7} we show simulation results with a power-law WTD~\eqref{eq:Psi-PL} of the concentrations versus time. In (a) we show a single  realization, while in  (b-c) we show ensemble averages. As expected, the relative fluctuations about the MF equilibrium point~(\ref{REPL}) decrease as we increase the number of realizations over which the concentrations are averaged.

\vspace{5mm}
\subsection*{Appendix D: Equilibrium for non-exponential WTDs}
\setcounter{equation}{0}
\renewcommand{\theequation}{D\arabic{equation}}
Here we obtain the equilibrium point in the general case of arbitrary $k_{\rm A}$, $k_{\rm B}$ and $k_{\rm C}$. Assuming a power-law WTD for the first reaction of \eqref{eq:reac} and using  memory kernel~(\ref{eq:thetaPLpprox}) valid for $\alpha_{\rm A}\!\gg\! 1$, the  equilibrium of \eqref{REPL} reads:
\begin{equation}\label{coex-pl_full}
\hspace{-3mm}\{a^*\!,b^*\!,c^*\}\!=\!\frac{\{(\alpha_{\rm A}\!-\!1)k_{\rm B},(\alpha_{\rm A}\!-\!1)k_{\rm C},(\alpha_{\rm A}\!-\!1/2)k_{\rm A}\}}{(\alpha_{\rm A}-1/2)k_{\rm A}+(\alpha_{\rm A}-1)(k_{\rm B}+k_{\rm C})}\!.
\end{equation}
Stability analysis shows that this fixed point remains a nonlinear center in the limit of $\alpha_{\rm A}\gg 1$, see Appendix A.

In the case of the gamma WTD, one can also compute the equilibrium point in the general case, for any $\alpha_{\rm A}$. Together with the relation $c^*=1-a^*-b^*$, we can solve the rate equations [Eqs.~(\ref{REPL})] for arbitrary  $\alpha_{\rm A}$, finding
\begin{equation}
\label{coex-gamma}
 \{a^*,b^*,c^*\} =\frac{\{2k_{\rm B}\,,2k_{\rm C}\,,k_{\rm A}\alpha_{\rm A} (3^{1/\alpha_{\rm A}}-1)\}}{2(k_{\rm B}+k_{\rm C})+k_{\rm A}\alpha_{\rm A}(3^{1/\alpha_{\rm A}}-1)}.
\end{equation}
The limit $\alpha_{\rm A}\to 1$ yields $(a^*,b^*,c^*)\!=\!(k_{\rm B},k_{\rm C},k_{\rm A})/(k_{\rm A}\!+\!k_{\rm B}\!+\!k_{\rm C})$. 
In contrast, for $\alpha_{\rm A}\gg 1$, $a^*\simeq k_{\rm B}/(k_{\rm B}+k_{\rm C}+(k_{\rm A}/2)\ln 3)$, $b^*\simeq k_{\rm C}/(k_{\rm B}+k_{\rm C}+(k_{\rm A}/2)\ln 3)$, and $c^*\simeq (k_{\rm A}/2)\ln 3/(k_{\rm B}+k_{\rm C}+(k_{\rm A}/2)\ln 3)$, yielding $a^*=b^*>c^*$
when $k_{\rm A}=k_{\rm B}=k_{\rm C}$. Another important limit is $\alpha_{\rm A}\to 0$.  To leading order in $\alpha_{\rm A}\ll 1$, the coexistence equilibrium becomes
$a^* \simeq 2k_{\rm B}/(k_{\rm A}\alpha_{\rm A})\,3^{-1/\alpha_{\rm A}}$, $b^*=(k_{\rm C}/k_{\rm B})a^*$ and $c^* \simeq 1-2
(k_{\rm B}\!+\!k_{\rm C})/(k_{\rm A}\alpha_{\rm A})\,3^{-1/\alpha_{\rm A}}$. While $a^*$ and $b^*$ are exponentially small,  $c^*$  asymptotically approaches 1. 

Finally, we consider more carefully the limit of  $\alpha_{\rm A}\to 1$. For this, we introduce $\epsilon\equiv \alpha_{\rm A}-1$.
Assuming that $|\epsilon|\ll 1$, to linear order in $\epsilon$, the memory kernel [Eq.~(\ref{memory_gamma})] becomes
\begin{equation}\label{thetagamapprox}
   \Theta_{\text{G}}(a,b,c)\simeq 1 -\left[(3/2)\ln 3-1\right]\varepsilon.
\end{equation}
Here, the coexistence equilibrium, \eqref{coex-gamma}, becomes
\begin{eqnarray}
\label{coex-gamma-approx}
 &&\hspace{-5mm}a^*=\frac{k_{\rm B}}{k_{\rm A}+k_{\rm B}+k_{\rm C}}\left[1+\left(\!\frac{3\ln{3}}{2}\!-\!1\!\right)\!\frac{\epsilon k_{\rm A}}{k_{\rm A}\!+\!k_{\rm B}\!+\!k_{\rm C}}\right]\!,\\
   &&\hspace{-5mm}c^*=\frac{k_{\rm A}}{k_{\rm A}+k_{\rm B}+k_{\rm C}}\left[1-\left(\!\frac{3\ln{3}}{2}\!-\!1\!\right)\!\frac{\epsilon(k_{\rm B}+k_{\rm C})}{k_{\rm A}\!+\!k_{\rm B}\!+\!k_{\rm C}}\right]\!,\nonumber
\end{eqnarray}
with $b^*=(k_{\rm C}/k_{\rm B})a^*$.
One can see that for a gamma WTD, changing $\alpha_{\rm A}$ to be below $1$, namely  $\varepsilon<0$, has the same qualitative effect as having a power-law WTD with finite (but large) $\alpha_{\rm A}$. Note that,  to obtain Eq.~(\ref{thetagamapprox}), we have plugged the leading-order in $\varepsilon$ equilibrium point [Eq.~(\ref{coex-gamma-approx})] into the subleading-order term  in  $\epsilon$ of $\Theta_{\text{G}}$. 

In the special case of $k_{\rm A}=k$ and $k_{\rm B}=k_{\rm C}=1$, Eq.~(\ref{coex-gamma-approx}) drastically simplifies
%
and we can find $k^*\!=\!k(\alpha_{\rm A})$  for which $a^*\!=\!b^*\!=\!c^*$:
    $k^*(\alpha_{\rm A}) = 1 + \left[(3/2)\ln 3 -1\right] (\alpha_{\rm A} - 1)$
[see Eq.~(\ref{kstar-full})]. The linear nature of the interface between the red and blue phases can be seen in Fig.~\ref{fig4}(d-e) close to $\alpha_{\rm A}=1$.

\end{document}